# How Should the Law Treat Future AI Systems? Fictional Legal Personhood versus Legal Identity

Heather J. Alexander*[1], Jonathan A. Simon*[1,2] and Frédéric Pinard[2]

[1]Laboratory for the Future of Citizenship
[2]University of Montreal
*lead authors, equal contribution

# Introduction

Today, artificial intelligence (AI) systems are more agentic – more autonomous and more unpredictable -- than ever. Recent advances in the field of general purpose AI, or "AI that can perform a wide variety of tasks,"[1] including agentic AI systems ("systems which can autonomously plan and act to achieve goals with little or no human oversight,"[2]) already pose risks to society.[3] A major, and underexplored, risk of agentic AI is the threat it poses to the coherence of laws and legal systems.

Most legal systems, including those of the U.S., Europe and Canada, draw a binary distinction between subjects of the law, who hold rights and duties, and objects of the law, which do not.[4] This distinction is fundamental to both the common and civil law systems, with roots dating back to the laws of ancient Rome and the early common law.[5] AI systems are currently classified as objects, not subjects, everywhere that laws have ruled on the matter (we survey law to this effect in §2 below)[6]

What happens when something that is not a subject (e.g a car, or a pet dog) acts like one? It depends on the area of law, but the norm is that effects caused by objects trace back to subjects. For example, what happens when an object causes harm that demands a legal remedy? Under modern law, either responsibility traces to a subject (via e.g. product liability frameworks for manufactured goods, *in rem* jurisdiction for suits against objects in relation to a person or persons, or strict, no-fault liability for owned animals), or the event is an "act of god."[7] Agentic AI systems, however, may be tough to fit into the subject/object dichotomy. The dichotomy works best where the objects in question are under the control of subjects, and predictable. But already today, AI systems often appear to act with purpose and intent, frequently in unpredictable ways. For this reason, some scholars argue that a moral or legal "responsibility gap" is already upon us, in cases of harms caused by today's AI systems, where this means a mismatch between how responsibility is attributed and how we think it ought to be.[8] Already some legislatures and judges struggle to make sense of the agentic nature of AI in torts and copyright law, pointing to even bigger and more general problems coming down the road.

---

[1] Yoshua Bengio, *The First International AI Safety Report: The International Scientific Report on the Safety of Advanced AI*, 2 SUPERINTELLIGENCE - ROBOTICS - SAFETY & ALIGNMENT 50 (2025). As lawyer Ryan Calo puts it, "(t)here is no straightforward, consensus definition of artificial intelligence," so this article will adopt the definition used by the AI Action Summit. Ryan Calo, Artificial Intelligence Policy: A Primer and Roadmap (Aug. 8, 2017), https://papers.ssrn.com/abstract=3015350.

[2] Bengio, *supra* note 1 at 8.

[3] *Id.* at 144–148; GREGORY C. ALLEN & GEORGIA ADAMSON, THE AI SAFETY INSTITUTE INTERNATIONAL NETWORK: NEXT STEPS AND RECOMMENDATIONS (2024), https://www.jstor.org/stable/resrep64561.

[4] In the words of J. Jaffe, "(t)here is no 'in-between' position of personhood for the purposes of establishing rights because entities are categorised in a simple, binary, 'all-or-nothing' fashion." *Matter of Nonhuman Rights Project, Inc. v Stanley (2015 NY Slip Op 31419) (Supreme Court, New York County 2015)*. See also SM. Solaiman, *Legal Personality of Robots, Corporations, Idols and Chimpanzees: A Quest for Legitimacy*, 25 ARTIFICIAL INTELLIGENCE AND LAW 155 (2017).

[5] For examples from history, see JOHN CHIPMAN GRAY, THE NATURE AND SOURCES OF THE LAW BY JOHN CHIPMAN GRAY 46–49 (1st Edition ed. 2019), https://doi.org/10.4324/9780429243417.

[6] For a philosophical discussion see DAVID J. GUNKEL, PERSON, THING, ROBOT: A MORAL AND LEGAL ONTOLOGY FOR THE 21ST CENTURY AND BEYOND (2023), https://mitpress.mit.edu/9780262546157/person-thing-robot/. For a legal discussion of the object-person distinction in the context of AI, see Chapter 3 of JAMES BOYLE, THE LINE: AI AND THE FUTURE OF PERSONHOOD (The MIT Press, 2024), https://mitpress.mit.edu/9780262049160/the-line/.

This looming tension between the object classification of AI systems and their apparent "agentivity" – their ability to convince people of things, create new images, and more generally act in the world in unpredictable ways, autonomously of the supervision of humans – is going to strain the *coherence* of many legal systems, leading to convoluted rulings and conflicts of laws. This tension, and the threat to legal coherence that it poses, will be our topic in this paper.

What do we mean by "coherence"? The rule of law is the bedrock of our legal and political order, the foundation of democratic governance and the guarantor of basic freedoms. Coherence is one of its essential conditions: the requirement that the law "make sense as a whole" — that its rules, principles, and decisions not be self-contradictory, arbitrary, or ad hoc.[9] Ronald Dworkin frames this as law's *integrity* — the idea that legal propositions should flow from a single, coherent set of principles that justify the system as a whole, i.e. that the law should "speak with one voice".[10]

---

[7] In cases of *in rem* jurisdiction, the action is against an object in relation to a person or persons, rather than directly against the person themselves. The classic use case is when a property or thing is identified but its owner is unknown, and a case against the unknown owner is sometimes styled as though against the objects themselves, e.g. as in United States v. 422 Casks of Wine, 26 U.S. 547 (1828) (1828). However, such cases do not ascribe legal responsibility to the casks of wine; they ascribe authority to the court to determine possessions of the casks. See Morris Cohn, *Jurisdiction in Actions in Rem and in Personam*, 14 WASHINGTON UNIVERSITY LAW REVIEW 170 (1929). In medieval times in Europe, objects and animals that caused harm were sometimes thought to contain evil spirits were granted subject status under the principle of *deodand*. Morris, cited above, notes that *deodand* may be the historical origin of *in rem* jurisdiction, but in modern law, *in rem* jurisdiction is simply a device to ensure a plaintiff can get relief in cases where the defendant is unreachable or unknown, not a means to recognize the agency or personhood of objects. See also Teresa Sutton, *The Deodand and Responsibility for Death*, 18 THE JOURNAL OF LEGAL HISTORY 44, 44–55 (1997).

[8] Andreas Matthias, *The Responsibility Gap: Ascribing Responsibility for the Actions of Learning Automata*, 6 ETHICS INF TECHNOL 175 (2004); David Nersessian & Ruben Mancha, *From Automation to Autonomy: Legal and Ethical Responsibility Gaps in Artificial Intelligence Innovation*, 27 MICHIGAN TECHNOLOGY LAW REVIEW 55 (2021); Nadia Banteka, *Artificially Intelligent Persons*, 58 HOUS. L. REV. 537 (2021); Vasudevan, *Addressing the Liability Gap in AI Accidents*, CENTRE FOR INTERNATIONAL GOVERNANCE INNOVATION (July 31, 2023), https://www.cigionline.org/publications/addressing-the-liability-gap-in-ai-accidents/.

[9] J. M. Balkin, *Understanding Legal Understanding: The Legal Subject and the Problem of Legal Coherence*, 103 THE YALE LAW JOURNAL 105, 114–115 (1993); Luc J. Wintgens, *Coherence of the Law*, 79 ARCHIV FÜR RECHTS- UND SOZIALPHILOSOPHIE / ARCHIVES FOR PHILOSOPHY OF LAW AND SOCIAL PHILOSOPHY 483 (1993), citing to MacCormick 1978, 125.

[10] See RONALD DWORKIN, LAW'S EMPIRE (1st edition ed. 1986), http://archive.org/details/EmpireDworkin; Julie Dickson, *Interpretation and Coherence in Legal Reasoning*, in THE STANFORD ENCYCLOPEDIA OF PHILOSOPHY

H.L.A. Hart, working in a different tradition, likewise treats consistency and predictability as central desiderata for a functioning legal system.[11] Joseph Raz, in his account of the rule of law, emphasizes that law must be capable of guiding the conduct of its subjects — a condition that is undermined when the law is riddled with contradictions or conceptual gaps.[12] Despite their deep disagreements on the foundations of law, these theorists all agree that coherence is essential to good law (Raz adds that it is essential to legitimate law; Dworkin adds that it is essential to count as law at all). Whether we think the law is "discovered" (as on realist or interpretivist accounts) or "created" (as in legal positivism), it is best understood as an integrated web or tapestry. On this view, threats to coherence are not minor doctrinal puzzles: they are threats to well-functioning (Hart) or indeed the very legitimacy (Raz) of the legal order. These threats manifest when precedent becomes convoluted and unpredictable, or when different areas of law conflict in ways that resist principled resolution — precisely the sorts of challenges that certain classifications of future AI systems are likely to provoke.

We will argue here that serious challenges to the coherence of law may arise from object classifications of some (but not all) future AI systems, as these become increasingly intelligent,[13] autonomous and human-like. Today's difficult copyright and liability cases may be manageable within an object framework, but they foreshadow far thornier issues to come. Consider, for example, smart autonomous robots,[14] (robots that are integrated with AI). These might serve as factory workers, as servants, or as companions. But we tend to get attached to our companions. What happens when people decide to marry their companion robots, or leave them an inheritance, or emancipate them? These are not things you can do with objects. Moreover, different jurisdictions may navigate these new challenges in different ways, leading to further inter-jurisdictional decoherence — conflicts of law that contribute to polarization and a lack of global consensus on how to integrate AI systems into our legal systems.[15]

---

(Edward N. Zalta ed., Winter 2016 ed. 2016), https://plato.stanford.edu/archives/win2016/entries/legal-reas-interpret/ (noting coherence as a distinctive requirement of legal reasoning), and Raymond Wacks, *Dworkin: The Moral Integrity of Law*, *in* PHILOSOPHY OF LAW: A VERY SHORT INTRODUCTION (Oxford Academic ed., 2nd Edition ed. 2014), https://doi.org/10.1093/actrade/9780199687008.003.0003.

[11] See H. L. A. (HERBERT LIONEL ADOLPHUS) HART, THE CONCEPT OF LAW 125–127 (New York : Oxford University Press ed. 1972), http://archive.org/details/conceptoflaw00hart (discussing legal consistency and predictability as core attributes of a functioning legal system); HANS KELSEN, GENERAL THEORY OF LAW AND STATE (Anders Wedberg trans., 1961) "Law is ... a set of rules having the kind of unity we understand by a system."

[12] See Joseph Raz, *Authority and Consent*, 67 VIRGINIA LAW REVIEW 103 (1981) (emphasizing law's role in guiding behavior as central to the rule-of-law ideal; cf. Joseph Raz, *The Relevance of Coherence*, 72 B. U. L. REV. 273, 273 (1992) (arguing coherence is necessary for legal clarity and obedience.)

[13] See for example *What Is Artificial General Intelligence ?*, GOOGLE CLOUD, https://cloud.google.com/discover/what-is-artificial-general-intelligence (last visited Sept. 8, 2025).

[14] The European Parliament, *European Parliament Resolution of 16 February 2017 with Recommendations to the Commission on Civil Law Rules on Robotics (2015/2103(INL))*, (2017), https://eur-lex.europa.eu/legal-content/EN/TXT/?uri=oj:JOC_2018_252_R_0026. See for example the robotics startup FigureAI at *Master Plan | Figure*, FIGUREAI (May 20, 2022), https://www.figure.ai/master-plan.

[15] Felix M. Wilke, *Dimensions of Coherence in EU Conflict-of-Law Rules*, 16 JOURNAL OF PRIVATE INTERNATIONAL LAW 163 (2020); A CONFLICT OF LAWS COMPANION 232 (Andrew Dickinson & Edwin Peel eds., Oxford Academic ed. 2021), https://doi.org/10.1093/oso/9780198868958.001.0001; Amol Shelar, *The Impact of AI on Conflict of Laws Artificial Intelligence in Conflict of Law: Navigating the Boon and Curse of Technological Justice*, 4 JUS CORPUS LAW JOURNAL 37 (2024).

How ought we respond to these challenges? The alternative to an object classification is to construe future AI systems – not all of them, but suitably advanced, suitably individuated ones – as persons, i.e., subjects of the law, rather than objects. Accordingly, there is a growing literature discussing the prospective "legal personhood" of future AI systems.[16] However, the topic is controversial,[17] the literature leaves key questions unresolved (beyond the obvious one of which systems or system-instances would qualify), and key legal distinctions are sometimes conflated.[18]

One common source of confusion in the AI personhood literature is the fact that there are two profoundly different (though intersecting) traditions in the law concerning persons, both in common law and civil law systems. There is one body of law for fictional persons, e.g. corporate persons, and another body of law for non-fictional persons, i.e. "natural" or physical persons, such as human beings.

**Fictional (Corporate) Legal Personhood:** What the civil law tradition calls "moral persons" (think "mores", i.e. conventions – the term `fictional' comes from the canon law notion, *persona ficta* and the latin *fingere,* meaning "made" rather than "unreal").[19] These are devices or conventions created to facilitate complex liability relationships between individuals.[20] The exact ontology has long been debated:[21] some say that corporations and other fictional persons are somehow "reducible to relations between the persons who own stock in them, manage them, and so forth."[22] Others maintain that fictional persons are organizational structures, i.e. bundles of "deontic facts" relating pre-existing individuals.[23] Crucially, the debate today is not whether fictional persons are fictional in the sense of being unreal, i.e. "imaginary" or "figments of lawyers' imaginations", but of how to understand their reality : there is no serious question that corporations exist, just as there is no serious question of whether money exists, or of whether e.g., the U.S. Constitution exists.[24]

---

[16] See for example Rafael Dean Brown, *Property Ownership and the Legal Personhood of Artificial Intelligence*, 30 INFORMATION & COMMUNICATIONS TECHNOLOGY LAW 208 (2021); Hon Katherine B Forrest, *The Ethics and Challenges of Legal Personhood for AI*, 133 THE YALE LAW JOURNAL 1176, 1231 (2024); Lawrence Solum, *Legal Personhood for Artificial Intelligences*, 70 NORTH CAROLINA LAW REVIEW 1231 (1992); Francis Rhys Ward, *Towards a Theory of AI Personhood*, in THE THIRTY-NINTH AAAI CONFERENCE ON ARTIFICIAL INTELLIGENCE (2024), https://openreview.net/forum?id=DOBYnbx1xr; Mark Fenwick & Stefan Wrbka, *AI and Legal Personhood*, in THE CAMBRIDGE HANDBOOK OF ARTIFICIAL INTELLIGENCE: GLOBAL PERSPECTIVES ON LAW AND ETHICS 288 (Cristina Poncibò, Larry A. DiMatteo, & Michel Cannarsa eds., 2022), https://www.cambridge.org/core/books/cambridge-handbook-of-artificial-intelligence/ai-and-legal-personhood/B2AC209680831AF553F966E05366B668; Visa A.J. Kurki, *The Legal Personhood of Artificial Intelligences*, in A THEORY OF LEGAL PERSONHOOD 175 (1 ed. 2019), https://academic.oup.com/book/35026/chapter/298856312; Brandeis Marshall, *No Legal Personhood for AI*, 4 PATTERNS (N Y) 100861 (2023).

[17] See for example *Europe Divided over Robot 'Personhood,'* POLITICO (Apr. 11, 2018), https://www.politico.eu/article/europe-divided-over-robot-ai-artificial-intelligence-personhood/.

[18] Brown, *supra* note 16 at 234.; Kurki, *supra* note 16. See generally Ward, *supra* note 16.; Fenwick and Wrbka, *supra* note 16; Solum, *supra* note 16.

[19] Maximilian Koessler, *The Person in Imagination or Persona Ficta of the Corporation*, 9 LOUISIANA LAW REVIEW 435 (1949).

[20] JOHN SEARLE, THE CONSTRUCTION OF SOCIAL REALITY (Free Press ed. 1995), https://epistemh.pbworks.com/f/6.+The+Construction+of+Social+Reality+(SCAN).pdf.

[21] John Dewey, *The Historic Background of Corporate Legal Personality*, 35 THE YALE LAW JOURNAL 655 (1926).

[22] Solum, *supra* note 16 at 1239.(Our italics.)

[23] See generally SEARLE, *supra* note 20.

The category of fictional legal persons has expanded over time. Rivers, for example, have emerged as special cases of fictional legal persons – but taxonomical care is required, because in such cases the person in question is not the river itself but an organization, existing in relation to a group of humans, with rights and duties pertaining to the river.[25] The river itself is not, strictly speaking, identical to the legal person, because a river cannot be stipulated into or out of existence by a legal body, while a fictional legal person can. In a slogan: fictional legal persons are *constituted* by law.

**Non-fictional (Natural) Legal Personhood**: Physical persons, such as human beings, whose legal rights as persons are based in underlying facts about their nature (e.g., their dignity). These persons exist independently of legal acts, though the recognition of some of their rights and duties (such as citizenship) may depend on such acts (such as birth registration). In civil law, these are called "*personnes physiques*." In a slogan: non-fictional legal persons are *recognized* by law.

Three crucial clarifications are necessary.

*First Clarification:* Both forms of personhood are forms of legal personhood. The key contrast is not between a legal form of personhood and some extra-legal form, but between two distinct forms of personhood before the law. Usage of "legal person" in common law can be confusing in this regard, as it sometimes refers only to corporate persons, but this is because one customarily uses the phrase "human" or "person" otherwise (and no laws codify the matter explicitly), not because strictly speaking human persons are not persons before the law -- *"person before the law"* is, of course, synonymous with "legal person".[26] In civil law the terminological ambiguity is avoided: there, it is generally codified or made explicit that there are two kinds of *personnes juridiques* – both *personnes morales* and *personnes physiques*.

---

[24] Admittedly, there is in philosophical ontology a form of rapacious reductionism according to which the only things that "really" exist are quantum fields permeating spacetime. On this extreme view, even we ourselves do not exist. Those prepared to question our own existence may of course also question the existence of corporations, money, and the like, but those who would deny the reality of corporations, money, etc. must tell us how the trick is done, if they do not propose to also question their own existence; a matter beyond the scope of sensible jurisprudence. See for discussion Barry Smith, *Towards an Ontology of Common Sense*, *in* THE BRITISH TRADITION IN TWENTIETH-CENTURY PHILOSOPHY 300 (Jaakko Hintikka ed., 1995), https://philarchive.org/rec/SMITAO-34.

[25] For an overview of fictional legal personhood as granted to rivers around the world, see Catherine J. Iorns Magallanes, *From Rights to Responsibilities Using Legal Personhood and Guardianship for Rivers*, 11 VICTORIA UNIVERSITY OF WELLINGTON LEGAL RESEARCH PAPERS 26 (2018) ; Erin L. O'Donnell & Julia Talbot-Jones, *Creating Legal Rights for Rivers: Lessons from Australia, New Zealand, and India*, 23 ECOLOGY AND SOCIETY 10 (2018).

[26] This is an example of what H.P. Grice (1975) calls "implicature": the pragmatic-linguistic phenomenon whereby an utterer can suggest something by an utterance that is not part of the literal meaning of the utterance, such as when one says "Cy ate some of the cherries" - this carries the implicature, but does not literally imply, that Cy did not eat all of the cherries. See Wayne Davis, *Implicature*, THE STANFORD ENCYCLOPEDIA OF PHILOSOPHY (Edward N. Zalta & Uri Nodelman eds., Spring 2024 ed. 2024), https://plato.stanford.edu/archives/spr2024/entries/implicature/.

Fortunately, there is a framework in international law for discussing non-fictional personhood as opposed to fictional personhood, in an unambiguous way, in contexts (such as the UN) where member states from both common and civil law systems work together and the applicable law is often blended. The key notion here is "legal identity", which amounts to the fact of being recognized as a non-fictional person. According to the United Nations operational definition,

> (l)egal identity is defined as the basic characteristics of an individual's identity. e.g. name, sex, place and date of birth conferred through registration and the issuance of a certificate by an authorized civil registration authority following the occurrence of birth. In the absence of birth registration, legal identity may be conferred by a legally-recognized identification authority. This system should be linked to the civil registration system to ensure a holistic approach to legal identity from birth to death. Legal identity is retired by the issuance of a death certificate by the civil registration authority upon registration of death.[27]

Establishing a legal identity for all is one of the UN's Sustainable Development Goals[28], the flagship global framework that provides the current organizing principle for the United Nations, and a version of legal identity is used by governments around the world to identify and register their populations. Under international law and the laws of most states, legal identity automatically provides access to basic rights like the right to life and due process, freedom from slavery, and freedom of conscience and speech.

_Second Clarification:_ These two categories of legal personhood are mutually exclusive and amount to very different things. This point can be missed in popular discussions of topics such as jurisdiction _in rem_, or the personhood of rivers (or other natural features), idols, animals, group rights or historical cases which may appear to be unsettled, such as slavery. But most of these cases belong to one category or the other, and the rest are exceptions that prove the rule. In jurisdiction _in rem_, a case is brought "against" an object, such as a ship or merchandise, as in _United States v. 422 Casks of Wine_ (1828)[29] or _United States v. 11 1/4 Dozen Packages of Articles Labeled in Part Mrs. Moffat's Shoo-Fly Powders for Drunkenness_ (1941).[30] But the law does not here actually give legal standing to ships, casks of wine, or shoo-fly powders. The point rather is to enable a court to assert jurisdiction over persons unknown or absent, e.g. to condemn or take possession of objects within the court's jurisdiction, whoever may own them. A ruling _in rem_ is binding on every party (person) with an interest in the relevant objects, as opposed merely to the parties litigant.[31]

Rivers and ecosystems are objects, sometimes associated with fictional legal persons – again, the person in question corresponds to a group of humans, with rights and duties pertaining to the river (see just above and §4 below). The motivating jurisprudential idea in such cases is not to expand the

---

moral sphere and e.g. bring rivers into the social and moral community (though see Stone 1972[32]), but rather to provide a more culturally appropriate legal structure than ownership or an easement to facilitate, e.g., land rights of indigenous groups (of humans). More generally, the rights of nature movement seeks to establish fictional legal persons corresponding to a range of natural features, mainly as a way to give indigenous or environmental groups standing to initiate legal actions concerning those features. Crucially, in such cases the creation of such a person involves designated special roles and duties for a class of specific people, allowing them to lend their will and agency to the common purpose expressed by the fictional person.

This can also be seen in cases involving idols (Solaiman 2017). Here some object of religious significance is granted personhood, but the substance of the legal act is to reorganize the rights and duties of existing persons designating them caretakers for the relevant object, giving them rights to sue on its behalf : hallmarks of fictional personhood, where in practice the new structure is a corporate one, requiring the will and agency of other, already existing people.[33]

Animals are considered objects in all jurisdictions with laws protecting them citing responsibilities between human or corporate persons (rather than responsibilities to the animals construed as subjects), even where the rationales for these laws concern the welfare of the animals in question. Groups such as PETA have made efforts to change the law, establishing some form of personhood for animals that would give activists standing to sue on behalf of their welfare. Were such efforts to succeed, there would be important open questions about the most suitable form of personhood to consider, but at present, the law universally considers animals to be objects, so there is no precedent here for the law embracing some hybrid of fictional and non-fictional forms of personhood. It bears stressing here that animal welfare laws exist, even though animals are classified as objects. In philosophical terms, we can say that the law already distinguishes between *patiency* (having a welfare) and *agency* (being a legal person).

As far as group rights are concerned, in law, these either refer to rights held by fictional legal persons, or rights held by individuals (non-fictional legal persons) in virtue of group membership. Slaves were a complicated hybrid category – treated like persons in some respects and objects in others – but this engendered extreme forms of decoherence (and ultimately war), illustrating the untenability of such arrangements, as we discuss in detail below (§3).[34]

Not only are the categories disjoint and fairly well-demarcated, but they entail very different things. Legal identity is the vehicle by which one accesses fundamental, non-derogable rights like the right to life, due process, freedom from slavery, and freedom of conscience. The ability of a state to revoke the legal identity of a non-fictional person is extremely circumscribed under international law,[35] and can only take place if there is a mechanism to ensure that the person retains some form of legal identity elsewhere (note that taking away someone's legal identity is a more drastic violation than denationalization, i.e. taking away someone's citizenship, though the latter is also heavily circumscribed by international law and, to be legal, requires that the person not be rendered stateless).[36] As the Office of the High Commissioner for Human Rights puts it, "Human rights are rights that every human being has by virtue of his or her human dignity."[37] In contrast, the rights of fictional persons are derogable (e.g. via *piercing the corporate veil*, where a court disregards a legal person to trace liability to involved individuals, or via disincorporation where a court dissolves a fictional person) and dependent on public policy goals. What the law brings into being by decree, it can take away.

---

[32] Christopher Stone, *Should Trees Have Standing*, 45 SOUTHERN CALIFORNIA LAW REVIEW 450 (1972).

[33] See the ruling of the Privy Council in Pramatha Nath Mullick vs Pradyumna Kumar Mullick on 28 April, 1925 (Bombay High Court 1925), and the discussion in Solaiman, *supra* note 4.

[34] Other historical examples of hybrid or hard to classify legal status also illustrate incoherence corrected in more recent law, including the status of children, women, indigenous persons, and outlaws. For example, "civil death" or *civiliter mortuus* was a punishment in which even basic rights such as the right to life were revoked. Modern doctrine of the inalienability of fundamental rights precludes this form of punishment (and the incoherences it engenders). See Ross Lerner, *Civil Death in Early Modern England*, 32 EXEMPLARIA 326 (2020).

Some discussions ask whether we should "give rights" to select AI systems. In light of the foregoing, we can see that this is ambiguous between two different meanings, one for each form of personhood: on the one hand, "giving rights" can mean *creating* a fictional legal person (e.g. incorporating a business partnership) and ascribing to it such rights and duties as we desire; on the other hand, "giving rights" can mean *recognizing* the non-fictional personhood of a physically existing individual, conferring its legal identity (a form of recognition of a pre-existing reality). This will be important below, because the rules for these two approaches to "giving rights" are very different. The deciding factor in which rights we "give" AI systems is pragmatic – a matter of what suits us best – if they are fictional legal persons. But if they are non-fictional legal persons then the question is what they deserve, based on the kinds of beings they already are, not what suits us.

*Third Clarification:* Neither form of personhood is by definition off-limits for AI systems. Though an AI system cannot be strictly identified with a fictional legal person (for the same reason that a river cannot), one might create fictional legal persons pertaining to AI systems, even those controlled by existing AI companies (much as one corporation might control another). At the same time, there are no contradictions in the idea that some AI systems could one day be recognized as non-fictional legal persons, despite their being "artificial". Care is required, since non-fictional legal persons are sometimes described as "natural". But "natural" here contrasts with "conventional", not with "manufactured" (the term "artificial" is sometimes used as synonymous with "fictional" but in such cases it means "conventional" rather than "manufactured"). Fictional persons are not "made" in the sense in which automobiles or robots and AI models are made; they are made in the sense in which contracts and declarations are made.[38] Of course, some jurisdictions may have laws, or *dicta*, that stipulate that natural persons are humans (and in one recent example, that AI systems are not persons[39]). But the underlying spirit of the distinction between fictional and non-fictional persons does not presuppose the humanity or indeed the biological substrate of non-fictional persons, and there are many jurisdictions where the matter is not settled, or could be revisited without undoing anything foundational (following roughly the procedures that we might follow if extraterrestrials sought refuge on our planet).

---

[35] See for example the discussion of legal identity in cases of armed conflict, Kathryn Hampton, *Born in the Twilight Zone: Birth Registration in Insurgent Areas*, 911 INTERNATIONAL REVIEW OF THE RED CROSS (2019), http://international-review.icrc.org/articles/born-twilight-zone-birth-registration-insurgent-areas. See also Heather Alexander, *Nomads and the Struggle for a Legal Identity*, 2 THE STATELESSNESS & CITIZENSHIP REVIEW 338 (2020).

[36] Laura van Waas, *Foreign Fighters and the Deprivation of Nationality: National Practices and International Law Implications*, *in* FOREIGN FIGHTERS UNDER INTERNATIONAL LAW AND BEYOND 469 (Andrea de Guttry, Francesca Capone, & Christophe Paulussen eds., 2016), https://link.springer.com/chapter/10.1007/978-94-6265-099-2_23.

[37] See for example Office of the United Nations High Commissioner for Human Rights (OHCHR), *What Are Human Rights?*, OHCHR, https://www.ohchr.org/en/what-are-human-rights (last visited Sept. 8, 2025).

[38] SEARLE, *supra* note 20.

Having clarified these points, it is clear that we face a stark choice regarding the legal personhood of future AI systems. We must choose between two different forms of legal personhood and an object framework (with its own subcategories: product, service, platform, etc). In doing so we must guard against the temptation to think of fictional (corporate) personhood as an easier, preliminary step in the direction of "premium" non-fictional personhood. There is no sense to be made of "upgrading" a fictional person into a non-fictional one. While corporations in some jurisdictions may have some of the same rights as humans, such as freedom of speech, the source of these rights is different, as are the rules relating to their durability and recognition. Corporations cannot turn into physical persons; physical persons cannot turn into corporations; and the laws governing them are radically different and derive from radically different traditions. Of course, humans (and other persons) can be shareholders in corporations (or other fictional persons); but in becoming a shareholder of a corporation one does not become a corporation. We must therefore treat fictional and non-fictional legal personhood for AI systems as fundamentally distinct, mutually exclusive statuses that might be associated with a given AI system (though conceivably different options might be best for different kinds of AI systems).

Our aim in this paper will be to consider these options through the lens of legal coherence : to assess whether we maximize overall long-term legal coherence by (A) maintaining an object classification for all future AI systems, (B) creating fictional legal persons associated with suitably advanced, individuated AI systems (giving these fictional legal persons derogable rights and duties associated with certified groups of existing persons, potentially including free speech, contract rights, and standing to sue "on behalf of" the AI system), or (C) recognizing non-fictional legal personhood through legal identity for suitably advanced, individuated AI systems (recognizing them as entities meriting legal standing with non-derogable rights which for the human case include life, due process, habeas corpus, freedom from slavery, and freedom of conscience). We will clarify the meaning and implications of each option along the way, considering liability, copyright, family law, fundamental rights, civil rights, citizenship, and AI safety regulation. We will tentatively find that the non-fictional personhood approach may be best from a coherence perspective, for at least some advanced AI systems. An object approach may prove untenable for sufficiently humanoid advanced systems, though we suggest that it is adequate for currently existing systems as of 2025. While fictional personhood would resolve some decoherence issues for future systems, it would create others and provide solutions that are neither durable nor fit for purpose. Moreover, our review will

---

suggest that "hybrid" approaches are likely to fail and lead to further incoherence : the choice between object, fictional person and non-fictional person is unavoidable.

This essay, a collaboration between a human rights lawyer trained in common and international law, a lawyer trained in civil law, and a philosopher, adds to the literature in part by attempting to better integrate the underlying philosophical questions within the framework of existing law and practice, both within common and civil law jurisdictions and in international law. The essay also adds to the literature on AI personhood by approaching the issue from a legal coherence perspective, instead of a perspective that makes moral assumptions about AI welfare. In particular, our argument is not of the form "future AI systems will be sentient and have moral status or welfare; therefore morality requires that we give them rights." Instead our argument is that recognizing some AI systems as non-fictional legal persons may maximize legal coherence, and legal policy should be guided by what maximizes legal coherence and upholding the rule of law – a point that, as we note above, scholars as disparate as Dworkin and Hart can agree on.

Here specifically is how we will proceed. In (§2), we assess the situation for current AI systems (at time of writing 2025). We observe that the classification of these AI systems as objects already causes some strain to the coherence of the legal system, by putting pressure on tort and copyright law, *inter alia*. However, we take the position that, concerning these systems, a subject classification (as either a fictional or a non-fictional legal person) would cause a greater strain on coherence than the *status quo* object classification does. In (§3), however, we argue that as AI systems develop to increasingly resemble humans both in capabilities and in appearance, their classification as objects will engender further, far more dramatic challenges to the coherence of the legal system – challenges for which, we argue, a subject classification for some suitably advanced systems will yield a more coherent system than if we remain with an object classification for all such cases. In (§4), we assess the pros and cons of the *fictional* legal personhood model for these (hypothetical) advanced systems, from a coherence perspective, an option with a basis in corporate law. In (§5), we assess the pros and cons of the *non-fictional* legal personhood model, i.e. the *legal identity* solution, deriving primarily from human rights and nationality law. In (§6) we evaluate, putting all of the pieces together. Here we conclude that all things considered, the best approach from a legal coherence perspective will maintain the object classification of a range of AI systems including current large language models, but will apply a non-fictional personhood classification to a restricted class of (hypothetical) advanced systems with sufficiently human-like capabilities. We note that the legal identity approach does not foreclose on AI systems being part of fictional legal persons in the future; indeed it facilitates that possibility, as fictional legal persons must ultimately be associated with groups of non-fictional legal persons.

## (2)The Object Classification of Current AI Systems : A Taste of Challenges to Come

Current (2025) law universally treats AI systems as objects—and therefore incapable of holding either rights or duties, being held responsible for harms, or speaking, or holding copyrights. This classification applies across torts and copyright cases currently before the courts, as well as in AI safety laws. This section assesses the extent to which this object classification is already raising problems of consistency and coherence in the legal system.[40] We examine three areas — torts, copyright, and AI safety laws — to argue two key points. First: current AI systems already impose genuine strains on legal coherence as limit cases of existing legal paradigms, defying straightforward resolution, in large part because they behave in subject-like ways — that is, in ways that only subjects have acted until now — but are not classified as subjects. Second: despite these strains, the challenges posed by current AI systems (e.g., chatbots like ChatGPT, Claude and Gemini, at the time of writing in 2025) do not yet warrant a shift from object to subject classifications. On balance, classifying today's systems as subjects would create more strain than maintaining the status quo. Accordingly, we concur with most scholars that current AI should not be classified as persons, even as we acknowledge the very real coherence challenges these systems present. Matters may differ for hypothetical future systems, a topic we address in later sections.

**Torts Law: The Responsibility Gap – AI can cause harms but is not responsible for them**

In torts law, AI systems universally remain objects incapable of legal responsibility, even as they demonstrate increasing autonomy and unpredictability. This creates what experts call the "responsibility gap"[41] — a conceptual strain that arises when highly autonomous objects cause harm but cannot be held liable.[42]

---

[40] The problems in classifying AI as an object, rather than as a person, under the law have been known since at least the 1990s. See for example Solum, note 16, 1231, 1233. See also RYAN ABBOTT, THE REASONABLE ROBOT: ARTIFICIAL INTELLIGENCE AND THE LAW (2020), https://www.cambridge.org/core/books/reasonable-robot/092E62F0087270F1ADD9F62160F23B5A.

[41] Lode Lauwaert & Ann-Katrien Oimann, *Moral Responsibility and Autonomous Technologies: Does AI Face a Responsibility Gap?, in* THE CAMBRIDGE HANDBOOK OF THE LAW, ETHICS AND POLICY OF ARTIFICIAL INTELLIGENCE 101, 105 (Nathalie A. Smuha ed., 2025), https://www.cambridge.org/core/books/cambridge-handbook-of-the-law-ethics-and-policy-of-artificial-intelligence/moral-responsibility-and-autonomous-technologies/B1A3D780C0C5364245198190AE134525.

[42] See for example Matthias, *supra* note 8; Jos Lehmann, Joost Breuker & Bob Brouwer, *Causation in AI and Law*, 12 ARTIF INTELL LAW 279 (2004); Monika Simmler, *Responsibility Gap or Responsibility Shift? The Attribution of Criminal Responsibility in Human–Machine Interaction*, 27 INFORMATION, COMMUNICATION & SOCIETY 1142 (2024); Michael Da Silva, *Responsibility Gaps*, 19 PHILOSOPHY COMPASS e70002 (2024). But see contra Joanna J. Bryson, *Robots Should Be Slaves*, 8 *in* NATURAL LANGUAGE PROCESSING 63 (Yorick Wilks ed., 2010), https://benjamins.com/catalog/nlp.8.11bry.

The fact that general-purpose AI systems can act in ways that were not explicitly programmed or intended by their developers or users raises questions about who should be held liable for resulting harm.[43]

The law's existing framework treats AI like any other object that can cause harm. Just as dogs remain legally classified as objects despite their capacity to injure,[44] AI systems cannot bear responsibility for damages they cause. Only legal subjects (i.e. persons) — such as owners, manufacturers, or users — can be held liable. This approach works adequately for predictable harms : an owner should know that their dog can bite and how to prevent it, and for this reason is understandably held liable when it happens.

Torts law allocates responsibility for harm based on well-established theories of causation, foreseeability and the recognition of a duty of care. Torts litigation functions most smoothly where responsibility can be directly and uncontroversially traced to one or multiple subjects, harm by harm, via these principles. However, torts law has also developed work-arounds, like the concept of strict liability, to enable restitution for plaintiffs when complicated technology and manufacturing processes makes causation and/or foreseeability difficult to establish.[45] Other concepts to fairly apportion liability in complex cases include shared liability and vicarious liability, whereby a principle is held liable for the actions of an agent, or employee.[46] These and other various principles of torts already developed to handle complicated cases (including mandatory insurance or government licensing and safety regulations) may prove suitable to manage the conceptual challenges that arise because of the various ways in which AI is complex and unpredictable (where dogs, in contrast, are in relevant ways predictable). However, this is by no means guaranteed or straightforward, and we must take things case by case.

*In re Toyota Motor Corporation* and *Benavides v. Tesla, Inc*[47] treat autonomous machines (in these cases, self-driving cars) as objects that are incapable of responsibility for their mistakes, instead allotting blame based on an analysis of the actions of the driver, the actions of the company responsible for the autonomous driving system and the actions of other actors along the supply chain (the car dealer, etc) The law also treats AI chatbots as objects that cannot speak under the law, instead allotting responsibility for harms caused by chatbot "speech" either to the

---

manufacturer or the user,[48] as demonstrated by ongoing litigation in *Garcia v. Character AI and al,* in Florida,[49] and *A.F. and al v. Character.AI and al*, in Texas.[50]

Because the chatbot cannot speak under the law, either the user or the chatbot company is responsible when the chatbot causes harm with its words.[51] The *Garcia* complaint alleges that it is the chatbot company that is responsible, even though the chatbots are designed to respond to the user's inputs in a spontaneous and natural-seeming way, meaning that the company does not have precise control over the chatbot's outputs.[52] The motion to dismiss in the *Garcia* case, filed by the chatbot company Character.AI, implied instead that it is the user of the chatbot who is responsible.[53]

Demonstrating the complexities of treating a speaking machine as an object that cannot "speak," the motion further asserted a 1st Amendment right on the part of the user to receive the chatbot's "speech," even if the chatbot, itself, was not speaking under the law. Yet it is unclear how these words, strung together by the chatbot, might be speech if they are not the chatbot's speech. The Court noted in its decision that "(d)efendants fail to articulate why words strung together by an LLM are speech,"[54] but the litigation on this question is pending. As an observer put it,

> (t)he "right-to-receive-speech" argument is critical because it allows the trial court to avoid the fascinating, technology-driven question of who or what actually is speaking—a chatbot or its creators? By concentrating on users' rights to receive speech, the court can dodge that issue.[55]

---

[48] Nina I Brown, *Bots Behaving Badly: A Products Liability Approach to Chatbot-Generated Defamation*, 3 JOURNAL OF FREE SPEECH LAW 389 (2023). See also Jan De Bruyne & Wannes Ooms, *Tort Liability and Artificial Intelligence: Some Challenges and (Regulatory) Responses*, *in* THE CAMBRIDGE HANDBOOK OF THE LAW, ETHICS AND POLICY OF ARTIFICIAL INTELLIGENCE 158, 163, 398 (Nathalie A. Smuha ed., 2025), https://www.cambridge.org/core/books/cambridge-handbook-of-the-law-ethics-and-policy-of-artificial-intelligence/tort-liability-and-artificial-intelligence/C0051569965DC5459FEF829AC88E2119. See also Jane R. Bambauer, *Negligent AI Speech: Some Thoughts About Duty*, 3 JOURNAL OF FREE SPEECH LAW (2023), https://papers.ssrn.com/abstract=4432822.

[49] Order on Motion to Dismiss - Garcia v. Character Technologies, Inc. | The Foundation for Individual Rights and Expression (United States District Court Middle District of Florida Orlando Division 2025).

[50] A.F., on Behalf of J.F. and al v. Character Technologies, Inc. and Al (UNITED STATES DISTRICT COURTEASTERN DISTRICT OF TEXASMARSHALL DIVISION 2024).

[51] For example, for the purposes of libel law. See the discussion in Eugene Volokh, *Large Libel Models? Liability for AI Output*, 3 J. FREE SPEECH L. 489 (2023). Volokh discusses liability for libel by a chatbot as involving only two possible parties, the user and the company. He does not even entertain the idea that the chatbot, itself, may be considered a third party. "Errors in what a company communicates can be defamatory regardless of whether the errors stem from direct human error in composing text or from human error in creating the technology that produces the text." *Id.*, 508, citing to Mark A. Lemley, Peter Henderson & Tatsunori Hashimoto, *Where's the Liability in Harmful AI Speech?*, 589, 620-25 (2023), https://papers.ssrn.com/abstract=4531029.

[52] Lemley, Henderson, and Hashimoto, *supra* note 49 at 91, 221, 350, 351.

[53] *Order on Motion to Dismiss - Garcia v. Character Technologies, Inc. | The Foundation for Individual Rights and Expression*.

[54] *Id.*

[55] Clay Calvert, *Free Speech or Culpable Conduct? When Role-Playing Chatbots Allegedly Harm Minors*, AMERICAN ENTERPRISE INSTITUTE - AEI (Feb. 19, 2025), https://www.aei.org/technology-and-innovation/free-speech-or-culpable-conduct-when-roleplaying-chatbots-allegedly-harm-minors/.

Yet, it is not clear that the question of whether a chatbot can "speak" can be wholly avoided in this litigation and, if the chatbot is not "speaking," it is not clear how to categorize and explain what, exactly, it is doing. In a related case, the motion to dismiss filed in *Waters v. Open AI* in the Northern District of Georgia argues that Open AI, a technology company that makes chatbots, is not liable for defamation because it is not responsible for statements made by its chatbot.[56]

More generally, traditional tort concepts strain under AI's complexity. Concepts like the industry standard for care, the foreseeability of a particular harm, or whether a manufacturer acted reasonably to prevent negligence, are hard to apply in cases when AI decision-making remains opaque (the "black-box problem"),[57] which it notoriously does, even when one has transparent access to a model's inner workings (this is the subject of the field of mechanistic interpretability research,[58]) and all the more so when one does not, which is often the case when model parameters involve proprietary information that companies resist disclosing.[59] Moreover, punitive damages, designed to deter future harms to protect society, may be difficult or impossible to establish.

> Long-standing punitive damages doctrine requires that the defendant *(the AI company or programmer)* act with malice or recklessness, which is unlikely to be provable in most cases of AI harm, in order to qualify for punitive damages.[60]

---

Importantly, these problems persist regardless of what kind of object AI is classified as, whether a product, service, license, and/or internet platform.[61] Even were courts to apply strict liability and label AI an "abnormally dangerous activity" under that torts doctrine,[62] or via the product liability framework, it may be impossible to identify and mitigate a design defect in the AI (a key element in those frameworks). This is, in effect, the "alignment problem", viewed from a liability perspective. AI is capable of "going rogue" -- acting in ways neither programmed nor intended by developers or users -- even when all available measures have been taken to prevent it from doing so, [63] sometimes surprising even expert AI engineers.[64] Aggravating the problem, AI technology advances so quickly that it can be difficult for scientists and experts to keep pace with its development, meaning that fixes for last year's models may not help with current ones.[65]

Note how this distinguishes AI from other dangerous technologies, like dynamite, for example. The latter is highly dangerous, but it causes harms within certain foreseeable and predictable parameters. The dangers of dynamite are well known, safety measures are possible, and harm can be mitigated.

In the face of these challenges, experts and policy makers have acknowledged the need for specially tailored rules for allocating responsibility for the harms caused by AI to existing subjects of the law (companies, users or society at large). As it was originally put in the (now withdrawn) Artificial intelligence Liability Directive,

> Current national liability rules, in particular based on fault, are not suited to handling liability claims for damage caused by AI-enabled products and services.[66]

It is important to note that a shift to a subject classification for AI systems has not been a part of any relevant proposals. Specific proposals include a government certification process, which would protect companies that are in compliance from most litigation, while imposing strict liability (where fault does not need to be proven) for harms caused by uncertified AI, as well as mandatory insurance.[67] The European Commission's aforementioned (now withdrawn) Artificial Intelligence Liability Directive proposed a combination of strict liability, a rebuttable presumption of causation, mandatory disclosure and insurance clauses to facilitate the establishment of liability for harms caused by AI on the part of tech companies.

---

[61] AI might be classified as a platform under 47 U.S. Code § 230. Note that experts find this classification unlikely. See for example Volokh, *supra* note 51.

[62] Weil, *supra* note 45.

[63] See for example INSTITUTE, *supra* note 57. See also Bryan Choi, *Negligence Liability for AI Developers*, LAWFARE (2024), https://www.lawfaremedia.org/article/negligence-liability-for-ai-developers.

[64] See for example Lindsey† and al., *supra* note 58.

[65] Bengio, *supra* note 1 at 177.

[66] EUROPEAN PARLIAMENT. DIRECTORATE GENERAL FOR PARLIAMENTARY RESEARCH SERVICES., *supra* note 59 at 9. Note that this proposal was withdrawn in early 2025.

[67] Matthew U. Scherer, *Regulating Artificial Intelligence Systems: Risks, Challenges, Competencies, and Strategies*, 29 HARVARD JOURNAL OF LAW & TECHNOLOGY 354 (2015). A regulatory certification process, such as a data hygiene certificate, for AI is also recommended by numerous agencies and research bodies, like the European Parliamentary Research Service (EPRS) in their 2020 report (which identified several key insights from the field of ethics and technology that seem to be relevant to AI, namely Transparency, bias, Socio-technical Systems, Responsibility, Risk Assessment, Ethical Technology Assessment and AI as a Social Experiment. See also also Sophia Falk & Aimee van Wynsberghe, *Challenging AI for Sustainability: What Ought It Mean?*, 4 AI ETHICS 1345 (2024).

Why do experts not consider subject classification as part of an envisioned package of specially tailored rules? We suggest the following explanation. Torts law developed in and was built for a world where no objects were as subject-like as current AI systems. But the strains on coherence here are incidental rather than essential. It is awkward to describe the outputs of a chatbot as anything other than "speech", but not incoherent. The real problem posed by the existence of AI in torts cases is not this sort of incoherence, but rather, the challenge they raise for legal policy of how to strike an appropriate balance between protecting subjects of potential harm, and facilitating innovation and growth. It is not obvious how a turn from an object classification to a subject classification would help with that goal, and there is reason to hope that a workable balance may be found, along the lines of solutions that experts are already proposing, while continuing to classify the relevant systems as objects rather than subjects.

To summarize: torts law confronts a very real strain because of the novel ways in which AI systems can cause sophisticated harms with the unpredictability and seeming autonomy of human subjects. Torts law functions most smoothly where objects that cause harms are easily predictable and controllable by humans, where apparent speech is produced by beings that are legally recognized as capable of speaking, and so on. However, torts law also has a rich repertoire of work-arounds for when things do not function smoothly. Admittedly, these work-arounds may in some cases produce less than ideal results – remedies that feel too demanding, or not demanding enough. Even so, if paired with new rules tailored to the case of generative AI, they may prove adequate to our needs without calling into question the object classification of AI systems, at least for now.[68]

## Copyright Law: The Creativity Paradox – AI Cannot Be Creative Under the Law, Yet It Can Make New Images and Text (or: "Do Transformers Transform?")

Copyright law faces a parallel coherence strain: AI systems can generate novel images and text, yet cannot hold copyrights, i.e. be legally recognized as author or creator of a work.[69] Only legal persons (fictional or non-fictional), as subjects of the law, can be legally recognized as authors or inventors, though recent and upcoming court decisions about the capacity of AI systems to "transform" work puts some strain on the justifications for this fact.[70] There is little historical precedent for cases where objects create things, and while animals evidence some of the hallmarks of creativity, their

---

[68] See Andrea Bertolini, *Robots as Products: The Case for a Realistic Analysis of Robotic Applications and Liability Rules*, 5 LAW INNOVATION AND TECHNOLOGY 214 (2013) for a related argument.

[69] C. E. Mammen and al., *Creativity, Artificial Intelligence, and the Requirement of Human Authors and Inventors in Copyright and Patent Law*, UNIVERSITY OF OXFORD (2024), https://ora.ox.ac.uk/objects/uuid:fe716d1e-e0bd-4b64-95b8-af149f040caa.

[70] Jozefien Vanherpe, *Artificial Intelligence and Intellectual Property Law*, in THE CAMBRIDGE HANDBOOK OF THE LAW, ETHICS AND POLICY OF ARTIFICIAL INTELLIGENCE 211 (Nathalie A. Smuha ed., 2025), https://www.cambridge.org/core/books/cambridge-handbook-of-the-law-ethics-and-policy-of-artificial-intelligence/artificial-intelligence-and-intellectual-property-law/7CEAEB5B630FCABF45DA9B0E9AAE0C3E. See also Legislative Services Branch, *Consolidated Federal Laws of Canada, Patent Act*, R.S.C., 1985, C. P-4 (2025), https://laws-lois.justice.gc.ca/eng/acts/p-4/FullText.html.

creativity is not recognized under the law, though we see hints of decoherence in the case law there as well.

In the famous "monkey selfie" case, for example, both the Wikimedia Foundation[71] and the Techdirt blog[72] took the position that Slater, the human photographer, could not hold copyright because the monkey, rather than Slater, took and hence created the picture, but that at the same time, since monkeys cannot hold copyrights (i.e., be creators), the image is in the public domain. The court did not accept this position, evading the threat of incoherence. However, in another animal law case, *Cetacean Community v. Bush*, we see a court required to disavow *dicta* from one of its own earlier rulings on a different matter (*Palila v. Hawaii*) in which it could not resist a punning remark suggesting that a bird had "winged its way into federal court as a plaintiff in its own right".[73]

At present, however, the U.S. Copyright Office is clear that only legal persons can hold copyright. The U.S. Patent Office also rejects patent applications listing an AI as the inventor.[74] Recently, the US Copyright Office took the extraordinary step of a preprint of their new guidance on the specific issue of generative AI; formal guidance is pending.[75]

One key question (in U.S. law) that remains unsettled for AI systems is whether such a system is capable of "transformative use" or "transformation" in using copyrighted works – whether (in the words of U.S. Supreme Court Justice Souter) the use in question "alters the original with new expression, meaning or message".[76] Work that makes use of copyrighted work but is not deemed transformative risks being a copyright violation under the provisions of the 1976 Copyright Act for "fair use". The matter is pressing, as most of today's generative AI systems make copious use of copyrighted work.

At present, policymakers contemplate, but have not yet enacted, fundamental changes to copyright law[77] and while courts have not yet handed down many decisions on this question, it is currently the subject of substantial litigation. The US Court in *Thomson Reuters v. Ross Intelligence,* a copyright infringement case, denied a motion for summary judgement by an AI company in February 2025, holding that the extent to which the AI was "transforming" the material it scraped to produce a new work or using existing work in a new way, rather than merely copying the material in violation of copyright, was a factual question for a jury.[78] *Thomson* indicates that courts may find that AI violates copyright because, as an object or tool, AI cannot transform anything. Yet, in a startling June 2025 order, the court in *Bartz and al v. Anthropic*, another ongoing "fair use" lawsuit, (in a remark meant as *dicta*) compared the learning process of Anthropic's AI "Claude" to that of a human author and suggested that its outputs may be transformative.[79]

---

[71] Louise Stewart, *Wikimedia Says When a Monkey Takes a Selfie, No One Owns It*, NEWSWEEK (Aug. 21, 2014), https://www.newsweek.com/lawyers-dispute-wikimedias-claims-about-monkey-selfie-copyright-265961.

[72] Mike Masnick, *Monkeys Don't Do Fair Use; News Agency Tells Techdirt To Remove Photos*, TECHDIRT (July 12, 2011), https://www.techdirt.com/2011/07/12/monkeys-dont-do-fair-use-news-agency-tells-techdirt-to-remove-photos/.

[73] See for example the "monkey selfie" case, Naruto v. Slater, No. 16-15469 (9th Cir. 2018) (UNITED STATES COURT OF APPEALSFOR THE NINTH CIRCUIT 2018); The Cetacean Community, Plaintiff-Appellant, v. George W. Bush, President of the United States of America; Donald H. Rumsfeld, United States of America Secretary of Defense, Defendants-Appellees, 386 F.3d 1169 (9th Cir. 2004) (U.S. Court of Appeals for the Ninth Circuit 2024); and the commentary in that ruling on Palila v. Hawaii Dept. of Land & Natural Resources, 471 F. Supp. 985 (D. Haw. 1979) 1106, 1107 (US District Court for the District of Hawaii 1979). The court wrote in *Palila* that the bird "has legal status and wings its way into federal court as a plaintiff in its own right." However, the same court in *Cetacean* makes a compelling case that its earlier remarks in *Palila* were meant as dicta and not intended to be precedential on the matter of animal standing.

[74] U.S. Register of Copyrights, *Copyright and Artificial Intelligence | U.S. Copyright Office*, (2025), https://www.copyright.gov/ai/.

[75] *Id.*

[76] Campbell v. Acuff-Rose Music, Inc., 510 U.S. 569 (1994) (UNITED STATES COURT OF APPEALS FOR THE SIXTH CIRCUIT 1994).

[77] See updates and guidance on the U.S. Register of Copyrights, *supra* note 74.

Just as we may ask how a monkey may simultaneously be author enough to create a new image, supplanting a human's claim to copyright, while not being author enough to claim copyright itself, so too we might wonder how an AI system might be author enough to transform material it has read, yet not be author enough to claim copyright. However, there are two points to be made here. First, the transformative use doctrine (developed by the U.S. Supreme court in a 1994 case concerning a *2LiveCrew* song[80]) can be understood as covering two different *kinds* of transformation – on the one hand, the kind in which a work with new content or meaning is created, but which draws on the original (as *2LiveCrew*'s song "Pretty Woman" parodies the Roy Orbison original), and on the other hand, the kind in which it is not the work itself that is a transformation, but rather, the way that the work is used. For example, a prior court held that legal publishing corporation Westlaw had the right to make copyrighted legal briefs available on their website, because the original authors of those briefs wrote them to help clients, while on Westlaw their use was instead to assist with research. Crucially, the court in *Bartz* cited this latter case, *White v. W. Pub. Corp,*[81] and its actual ruling concerned the AI model Claude itself as the product in question, rather than Claude's outputs. However, as mentioned above, the ruling in *Bartz* goes on to offer *dicta* suggesting that Claude's outputs might well be transformative in the former, "new creation" sense.[82]

The future is unclear for this line of "fair use" litigation regarding the ability of AI to transform existing work into something new. There is unlikely to be a perfect solution that comprehensively protects human creators while also fueling unlimited growth and development in (data-hungry) AI. In effect, we have three options : (a) ban AI-outputted content, holding that it is not copyrightable and not fair use, even when there is no commercial intent and the work would pass the transformation test if written by a human, (b) allow that AI content can be fair use but render it public commons, i.e. not copyrightable, or (c) allow that AI content (or AI-assisted content) be copyrightable.

---

[78] See for example Thomson Reuters Enterprise Centre GmbH and al v. ROSS Intelligence Inc. | District of Delaware | United States District Court (THE UNITED STATES DISTRICT COURT FOR THE DISTRICT OF DELAWARE 2025). In Canada, see for example *CanLII v. Caseway: Future of AI-Generated Legal Databases on Trial | IP Law 422 001*, https://iplaw.allard.ubc.ca/2024/11/09/canlii-v-caseway-future-of-ai-generated-legal-databases-on-trial/ (last visited Sept. 9, 2025); A list of the copyright lawsuits in the United States is available at https://www.wired.com/story/ai-copyright-case-tracker/. In November 2024, major Canadian news organizations – including *The Canadian Press*, *Torstar*, *The Globe and Mail*, *Postmedia*, and *CBC/Radio-Canada* – filed a lawsuit against OpenAI. The publishers alleged that OpenAI used their copyrighted content without authorization to train its AI models, such as ChatGPT. As of May 2025, the case is ongoing, highlighting the tension between AI development practices and intellectual property rights.
[79] Note that the case will go to trial regarding Anthropic's use of pirated material to train Claude. Bartz v. Anthropic PBC - Order on Fair Use (United States District Court, Northern District of California 2025).
[80] *Campbell v. Acuff-Rose Music, Inc., 510 U.S. 569 (1994)*.
[81] *Id.* at 13, citing White and al v. West Publishing Corporation and al, No. 1:2012cv01340 - Document 79 (S.D.N.Y. 2013) (UNITED STATES DISTRICT COURT SOUTHERN DISTRICT OF NEW YORK 2013).
[82] *Campbell v. Acuff-Rose Music, Inc., 510 U.S. 569 (1994)*; *Bartz v. Anthropic PBC - Order on Fair Use*.

From a public policy standpoint each option has costs. Banning AI content in effect outlaws generative AI, a move with possible economic downsides, if we consider the current boom in this technology to be a good thing. Allowing that Claude's outputs are fair use (when they differ just enough from existing works) risks allowing AI systems to undercut existing human creators, offering up just-different-enough works for free to read instead of the originals on which Claude has trained. Allowing copyrights for such AI-generated works risks suffocating future creators, by saturating the marketplace with "new" works at a potentially rapid pace, taking over the space of new ideas left open for human creators to fill.

From the standpoint of object vs. subject classification, care is required on any route we choose, but if such care is exercised, coherence may be preserved, even on an object classification. On either of the approaches that deny copyright claims for AI outputs, an object classification is clearly the more coherent – if the systems in question are legal persons, capable of creation, then denying their copyright claims would seem capricious.

On such approaches, where we deny copyright claims, we must exercise care in how we rule on the issue of transformation (we note the irony that most generative AI systems belong to a family of algorithms known as "transformers"[83]). If we opt for option (a) – denying fair use as well as copyright – then we must either modify fair use doctrine by adding a fifth factor (e.g. that AI not be involved) or by refining doctrine on creative transformation, for example, by requiring *both* that it involve new expression or message, and also that this be a message intended by a legal person. While this approach may draw the ire of some literary critics who maintain the death of the author,[84] for the purposes of copyright law it stands to reason that we take the intent of the author, understood as a (legal) subject, into account.

If we opt for option (b) – allowing fair use while denying copyright – then we must refine doctrine in a different way. Here there is some risk of incoherence, if it comes to cases where someone makes commercial use of AI outputs that we take to pass the creative transformation test, while denying copyrightability. If it is creative enough to pass the creative transformation test, why is it not creative enough to be copyrighted? However, proper clarification of rulings (and in future legislation) can ensure coherence, by clarifying that the creative transformation test is *not* meant to take the intent (or indeed existence as legal person) of the author into account, but merely whether readers can *derive* (or *"receive"*) new expression, meaning or message from the work (c.f. the discussion of "right to receive speech" above). In effect, options (a) and (b) can both be coherent, but coherence requires that courts (and legislators) decide whether they agree with Roland Barthes' claims about the death of the author (option b) or oppose them (option a), at least as far as AI outputs are concerned.

On option (c), where we allow copyright claims for AI-generated or assisted work, there is again no tension with an object classification, as long as we are clear that the copyrights go to the humans who make or use the relevant AI systems, and not to the AI systems themselves. If we want to give AI systems themselves the ability to hold copyright, then of course it makes more sense for them to be legal persons, but that is not a desideratum on the table in current cases. We provisionally conclude that while there are threats of incoherence to watch out for in copyright law if rulings are not explained and clarified carefully, a move from a subject to an object classification is not the obvious solution at this time.

**AI Safety Regulation: Regulating Autonomous Objects That Can "Go Rogue"**

In addition to existing legal frameworks like torts and copyright, AI is the subject of numerous proposed, specialised and bespoke laws to regulate its safety and use. In fact, much of the AI safety debate is specifically concerned with the need for specialized laws to address the dangers posed by rogue, or autonomous, AI. Despite this risk owing to the subject-like autonomy of such systems, a survey of existing laws and policies to regulate AI demonstrates almost universal agreement that relevant AI systems are objects, not subjects, under the law, and therefore any of the harms that such systems cause must be attributed to some other subject of the law (a human or legal person (company)), including in cases where an AI "goes rogue."

Proposed and current laws and policies to regulate AI include, for example, Utah's S. B. 149, Ohio's H. B. 469, Idaho's § 5-346,[85] US federal law,[86] President Biden's (rescinded) and President Trump's Executive Orders on AI,[87] the European Union's AI Act (the "EU AI Act", one of the first comprehensive legislation regulating AI),[88] Canada's Artificial Intelligence and Data Act (AIDA) (withdrawn),[89] the National Institute of Standards and Technology (NIST) AI Risk Management Framework,[90] the Japanese AI Guidelines for Business Ver1.0,[91] and the OECD AI Principles.[92]

---

[85] See also Kirk A. Cullimore, *supra* note 39; Walt Brooks, *supra* note 39. Note that the Utah law uses "legal person" in the comprehensive sense covering both fictional, corporate personality and non-fictional, physical personality.

[86] Such as *2021 U.S. Code :: Title 15 - Commerce and Trade :: Chapter 119 - National Artificial Intelligence Initiative :: Sec. 9401 - Definitions*, Justia Law, https://law.justia.com/codes/us/title-15/chapter-119/sec-9401/ (last visited Sept. 9, 2025). See also *The 2025 AI Index Report | Stanford HAI*, Chapter 7, https://hai.stanford.edu/ai-index/2025-ai-index-report (last visited Sept. 9, 2025).

[87] The White House, *Executive Order on the Safe, Secure, and Trustworthy Development and Use of Artificial Intelligence*, The White House (Oct. 30, 2023), https://bidenwhitehouse.archives.gov/briefing-room/presidential-actions/2023/10/30/executive-order-on-the-safe-secure-and-trustworthy-development-and-use-of-artificial-intelligence/; The White House, *Statement by President Biden on the Executive Order on Advancing U.S. Leadership in Artificial Intelligence Infrastructure*, The White House (Jan. 14, 2025), https://bidenwhitehouse.archives.gov/briefing-room/statements-releases/2025/01/14/statement-by-president-biden-on-the-executive-order-on-advancing-u-s-leadership-in-artificial-intelligence-infrastructure/.

Other proposed and model regulations and declarations also treat AI as an object, such as those of the European Artificial Intelligence Board, the Council of Europe Framework Convention on AI, the UK's AI Opportunities Action Plan, and the UN's AI Advisory Body. The Vatican has also released guidelines on AI that classify it as an object.[93] The Texas Attorney General has launched an investigation into AI that may pose a risk to children, classifying AI as a potentially harmful product, which is a kind of object.[94]

---

[88] See the EU AI Act, *supra* note 59. As per its Article 1 (Subject Matter'), the EU AI Act aims to improve the internal market's functioning (bounded by the borders of the European Union) and promote the development and uptake of human-centric and trustworthy AI while ensuring a high level of protection against the harmful effects of AI systems in the European Union (the "EU"). It establishes harmonized rules for placing AI systems on the market, putting them into service, and their use within the EU across Member States, and also lists prohibited AI practices, among others.

[89] *Government Bill (House of Commons) C-27 (44-1) - First Reading - Digital Charter Implementation Act, 2022 - Parliament of Canada*, (2022), https://www.parl.ca/documentviewer/en/44-1/bill/C-27/first-reading. The AIDA was halted amidst a prorogation of the Canadian Parliament in January 2025. See also Blair Attard-Frost, *The Death of Canada's Artificial Intelligence and Data Act: What Happened, and What's Next for AI Regulation in Canada? | Montreal AI Ethics Institute*, (Jan. 17, 2025), https://montrealethics.ai/the-death-of-canadas-artificial-intelligence-and-data-act-what-happened-and-whats-next-for-ai-regulation-in-canada/; Treasury Board of Canada Secretariat, *Guide on the Scope of the Directive on Automated Decision-Making*, (June 12, 2025), https://www.canada.ca/en/government/system/digital-government/digital-government-innovations/responsible-use-ai/guide-scope-directive-automated-decision-making.html, which enforces "departments that are using automated decision systems [developed or procured after April 2020] to fully or partially automate an administrative decision" – including those that rely on AI – to be "transparent, accountable and fair in automated decision-making".

[90] On January 26, 2023, NIST has developed the *NIST AI Risk Management Framework (AI RMF)* to better address risks to individuals, organizations, and society related with artificial intelligence (AI). The framework is designed for voluntary use and to enhance the ability to "incorporate trustworthiness considerations into the design, development, use, and evaluation of AI products, services, and systems". See Gina M. Raimondo, Secretary, Artificial Intelligence Risk Management Framework (AI RMF 1.0) (2023), http://nvlpubs.nist.gov/nistpubs/ai/NIST.AI.100-1.pdf.

[91] Ministry of Internal Affairs and Communications Ministry of Economy, Trade and Industry, AI Guidelines for Business Ver 1.0 Compiled (2024), https://www.meti.go.jp/english/press/2024/0419_002.html. Although not mandatory, the Guidelines propose specific desirable approaches for AI-related businesses entities in Japan. See also Tomomi FujikougeNaoto KosugeFumiya Matsumoto, *Understanding AI Regulations in Japan - Current Status and Future Prospects | DLA Piper*, https://www.dlapiper.com/en-ca/insights/publications/2024/10/understanding-ai-regulations-in-japan-current-status-and-future-prospects (last visited Sept. 9, 2025).

[92] OECD AI Principles (2019) (updated in 2024).

[93] Vatican City State, Pontifical Commission for State of Vatican City, Guidelines on Artificial Intelligence (2025), https://www.vaticanstate.va/images/N.%20DCII.pdf (available in Italian only).

[94] Ken Paxton, *Attorney General Ken Paxton Launches Investigations into Character.AI, Reddit, Instagram, Discord, and Other Companies over Children's Privacy and Safety Practices as Texas Leads the Nation in Data Privacy Enforcement | Office of the Attorney General*, texasattorneygeneral (Dec. 12, 2024), https://www.texasattorneygeneral.gov/news/releases/attorney-general-ken-paxton-launches-investigations-characterai-reddit-instagram-discord-and-other.

Some of these laws and proposed regulations, however, are already internally at risk of incoherence in that they simultaneously treat AI as something that will make decisions, have agency, show intent and act independently of humans, but also as an object. For example, Colorado's SB 205 of 2024, which is scheduled to take effect in 2026, defines a "high-risk" AI system as a system that "makes, or is a substantial factor in making, a consequential decision." Yet, all decisions made by AI, no matter how autonomous, must be attributed to a human actor or an AI company.

The converse, where we try to subsume AI wrongdoing under actions taken by humans, also gives rise to problems. Consider, for example, the recently passed Texas H.B. No.149 states in Sec. 552.052.,

> MANIPULATION OF HUMAN BEHAVIOR. A person may not develop or deploy an artificial intelligence system in a manner that intentionally aims to incite or encourage a person to: (1) commit physical self-harm, including suicide; (2) harm another person; or (3) engage in criminal activity.[95]

Of course, cases where an AI manufacturer deliberately sets out to cause harm should be legislated against. But this section appears intended to address harms caused by AI systems in general. By focusing on the harms that humans themselves might commit, this article fails to capture the various ways in which AIs may produce harmful content even if not explicitly trained to do so (i.e., going rogue).

Here, as in the cases of tort and copyright, we find that a genuine strain to coherence arises because our legal framework is not accustomed to objects that behave to this extent like subjects – but the problem is more one of requiring us to exercise care in refining our legal structures to accommodate this possibility, than of requiring us to shift from an object to a subject classification, which in the case of safety law does not clearly address the actual underlying problem, and may engender further complications of its own, a point on which we elaborate just below.

**Why Legal Personhood for Current Systems Would Create Greater Incoherence**

While the preceding analysis demonstrates real strains in maintaining AI's object status, reclassifying current AI systems as legal persons would potentially generate more severe coherence problems and practical difficulties than it would resolve. For today's AI systems, the complications of personhood outweigh those of maintaining object classification, as far as torts, copyright and safety laws are concerned. We have already explained above why a subject approach is not needed for such cases; here we identify further problems that a subject approach would create.

First, granting legal personhood to current AI would create significant problems of legal accountability. Legal persons must be capable of bearing both rights and duties—entering contracts, owning property, and facing meaningful legal sanctions. On the non-fictional model of personhood, there is much theoretical work left to be done to identify what such a notion (of meaningful legal sanction) might be, for a system like ChatGPT that is an abstract mathematical model, implemented by a different server every time a new instance is called.

---

[95] *AN ACT Relating to Regulation of the Use of Artificial Intelligence Systems in This State; Providing Civil Penalties (H.B. No. 149)*, H.B. NO. 149 (2025), https://capitol.texas.gov/BillLookup/Text.aspx?LegSess=89R&Bill=HB149.

Legal persons require clear identification, registration, and boundaries for basic legal operations— serving process, appearing in court, maintaining continuous identity for contractual obligations. Yet AI systems can be copied, versioned, fine-tuned, merged, or deleted. If a language model is copied and deployed in multiple instances, which instance bears the legal personhood? When a model is retrained or updated, does it remain the same legal person or become a new one? If two AI systems are merged, whose legal obligations survive?

On the fictional personhood model, meaningful sanction can be understood as per standard corporate law (and the vagaries of the "identity" of the AI system can be finessed, because there is not a strict identity between the AI system and the fictional legal person). Here the issue is that there arguably already are enough legal persons involved – the AI companies themselves. Moreover, companies such as OpenAI and Anthropic are already free to create and control other corporate entities. The question is how this move would represent any advance or resolution of the novel challenges discussed above.  Transferring liability to a corporate subsidiary does not bridge the responsibility gap in cases where we intuit that the manufacturer is not liable, nor solve the social problem of protecting artists from supplantation by AI.  And construing AI safety law as regulating a special class of corporations controlled by AI companies is little different from the current solution, on which we think of these regulations as governing AI companies. Carving out some entirely new subtype of fictional legal person (as has been done for rivers) might help us overcome some of these obstacles, but then we must confront anew the question of what meaningful sanction can amount to for this new form of corporate person.

Second, AI personhood for current systems would raise serious concerns about fairness in distributing benefits and burdens. Consider copyright: if AI could hold copyrights as legal persons, a single company might spawn hundreds of AI "persons," each generating copyrightable works at unprecedented speed. Unlike human creators who produce works at human pace over human lifespans, AI persons could potentially accumulate vast intellectual property portfolios, fundamentally altering market dynamics in ways our copyright system – premised on human creative capacity – never anticipated.[96] Crucially,though, as we note above, this problem would arise if we allow humans to hold copyrights for works they made in collaboration with AI systems, so the problem here is not purely a question of subject vs object classification.

In torts, AI persons without independent assets would be effectively judgment-proof.[97] Legal personality for AI/robots might in the worst case remove the checks placed on AI companies by the courts, shielding them from responsibility for harms (thus safeguarding corporations, manufacturers and users from accountability) and removing the incentives to make AI safe.[98] Mariano-Florentino Cuéllar, former justice of the Supreme Court of California, noted that, "it would be difficult to justify treating someone as essentially immune from whatever culpability they would otherwise have simply because he or she uses a robot or relies on an AI system."[99] While humans and companies profiting from AI's actions might escape liability, injured parties could be left pursuing empty remedies against assetless AI defendants. Admittedly, personhood for AI could also be coupled with safeguards (*e.g.*, mandatory insurance or human oversight) to prevent it from becoming a liability loophole,[100] but these distributional effects would require substantial restructuring of existing legal frameworks.[101] To be clear, many of these issues will remain pressing even for the sort of future AI systems whose personhood we think should be taken more seriously in the near future, especially if the personhood of these is construed on the fictional model, an issue we return to in sections four and five below.

---

[96] See on emerging copyright litigation and AI market-scale risks Katrina Geddes, *AI Liability for Intellectual Property Harms*, LAWFARE (2024), https://www.lawfaremedia.org/article/ai-liability-for-intellectual-property-harms.

[97] See Andrew D Selbst, *NEGLIGENCE AND AI'S HUMAN USERS*, 100 BOSTON UNIVERSITY LAW REVIEW 63 (2020). Andrew D. Selbst notes that decision-assistance AI often produces statistically unpredictable errors - "where the unforeseeable error is the rule, not the exception" (p. 1322) - causing negligence doctrines to collapse and leaving victims unable to establish fault when no human actor is identifiable or at fault. See also Joanna Bryson, Mihailis Diamantis & Thomas D. Grant, *Of, for, and by the People: The Legal Lacuna of Synthetic*

**Conclusion: Coherent for Now, But Straining**

Across torts, copyright, and safety regulation, current law maintains AI's object status despite mounting conceptual strains. The responsibility gap, creativity paradox, and contradictions in safety regulation all stem from the same source: AI systems that exhibit subject-like qualities—autonomy, unpredictability, apparent creativity—while remaining legal objects.

For current AI systems, the balance of complications favors maintaining object classification for now while adapting our frameworks to address their unique characteristics. The strains this creates, while real, remain more manageable than those that would arise from premature personhood. The coherence problems noted so far, while genuine, can be addressed through targeted legal innovations: strict liability, mandatory insurance, modified copyright frameworks, and safety regulations that, however awkwardly, attribute liability for AI decisions to human actors. There is also a growing body of research on AI welfare, which suggests that some AI systems, even current ones, may have mental states of their own, and possibly even be capable of suffering. However, as is the

---

*Persons*, 25 ARTIFICIAL INTELLIGENCE & LAW 273 (2017), who notably invoke challenges of coherence against the proposal of AI personhood, though they conclude that it is to be avoided in all cases not only current ones, where we reach a different conclusion below.

[98] Banteka, *supra* note 8; Bryson, Diamantis, and Grant, *supra* note 97.

[99] *Reconciling Law, Ethics, and Artificial Intelligence: The Difficult Work Ahead | Stanford HAI*, https://hai.stanford.edu/news/reconciling-law-ethics-and-artificial-intelligence-difficult-work-ahead (last visited Sept. 9, 2025).

[100] Banteka, *supra* note 8 at 596.

[101] See the EU proposal warning against legal fragmentation and the need to reform its liability framework in order to introduce new rules specific to damage caused by AI systems, "The new rules intend to ensure that persons harmed by AI systems enjoy the same level of protection as people harmed by other technologies in the EU", at Tambiama Madiega, *Artificial Intelligence Liability Directive*, (2023), https://www.europarl.europa.eu/RegData/etudes/BRIE/2023/739342/EPRS_BRI%282023%29739342_EN.pdf.

case for animal welfare, this can be addressed within extant legal frameworks classifying relevant systems as objects.[102]

That being said, as this section demonstrates, current AI already pushes against the boundaries of object classification. The challenges discussed here may prove minor compared to those posed by truly autonomous future systems—a prospect we examine in subsequent sections. As AI technology evolves toward greater autonomy, embodiment, and social integration, the calculus presented here may shift, making subject classification the more coherent choice. We differ from other critics of AI personhood in that, while we agree that AI personhood is a problematic idea for current systems, we expect that this will change as the technology develops further.

# (3) Future AI Systems' Object Status will Further Strain Coherence

An object classification for future AI systems risks serious coherence problems far beyond the strains we see in torts, copyright and AI safety laws today, as Alexander (2023) and others have pointed out.[103] Saudi Arabia already granted "citizenship" to a robot in 2017, sparking enormous debate,[104] while in 2024, the state of Utah passed a blanket law denying the personhood of robots,[105] including their possible future citizenship, pointing towards the possibility of future conflicts between jurisdiction over the personhood, rights and citizenship of AI systems and robots. This section will explore some of the areas of law that will be impacted, showing why object classifications for future systems will pose even more profound and systematic risks of coherence than they do for current systems.

**Family Law**

Family law, which is not currently relevant for AI, may become so. With humans already forming relationships with AI, there is the very real possibility that humans in the future will want to, for example, marry their companion robots (which will likely implement sophisticated AI systems). Laws allowing humans to marry their companion robots while maintaining the status of companion robots as objects, however, will create severe coherence problems. Additionally, some jurisdictions may allow humans to marry their robots, while others may refuse to recognize, or even ban, such marriages, creating a conflict of laws. Experts have already noted the problem of conflict of laws from agentic AI[106] and the growing need for harmonization at the regional and international levels.[107]

---

[102] Robert Long and al., Taking AI Welfare Seriously (Nov. 4, 2024), http://arxiv.org/abs/2411.00986; Patrick Butlin and al., Consciousness in Artificial Intelligence: Insights from the Science of Consciousness (Aug. 22, 2023), http://arxiv.org/abs/2308.08708.

[103] Heather Alexander, *COMMENT: The United Nations and Robot Rights*, 20 CANADIAN JOURNAL OF LAW AND TECHNOLOGY 257 (2022). See also Robert Booth & Robert Booth UK technology editor, *AI Could Cause 'Social Ruptures' between People Who Disagree on Its Sentience*, THE GUARDIAN, Nov. 17, 2024, https://www.theguardian.com/technology/2024/nov/17/ai-could-cause-social-ruptures-between-people-who-disagree-on-its-sentience.

[104] While the government of Saudi Arabia granted citizenship to a robot named Sophia in 2017, it is unclear what legal effect this has had. See also *Should Robots Be Citizens? | British Council*, https://www.britishcouncil.org/anyone-anywhere/explore/digital-identities/robots-citizens (last visited Sept. 9, 2025).

[105] Walt Brooks, *supra* note 39.

Why might allowing humans to marry their companion robots cause coherence issues? Objects cannot legally sign contracts, including marriage contracts, even if they are physically able to write names on a page. Only subjects, and not objects, can give their consent, i.e. "express their will," a core requirement for a legal marriage in most jurisdictions. An exception might be made under the marriage laws of a particular country whereby a limited class of objects, such as companion robots, can marry, but this will create decoherence with other areas of law, such as adoption, inheritance, divorce, tax law and more. Objects do not pay taxes, so married couples where one spouse is a robot might be unable to take advantage of favorable tax laws. The robot spouse would have no legal rights towards adopted children, meaning they might not be able to take the children on a trip, or even sign them up for a school activity, if the human spouse is not present, undermining a core principle of marriage.[108]

Conflicts of law may also develop, as some jurisdictions ban human-robot marriage. While the law currently bans humans from marrying objects, animals and children for reasons of public policy, there has been little organized push-back against these laws. Banning robot-human marriage entirely, however, might come to be seen by the public in some countries as violating the right of humans to freely choose their spouse. The Universal Declaration on Human Rights and the European Convention on Human Rights, to take but two examples, affirm the right of all humans to marry without discrimination.[109] The right to choose one's spouse has also been upheld by national courts and laws such as the US Supreme Court in *Obergefell v. Hodges,*[110] Article 21 of The Constitution of India, and Canada's Civil Marriage Act. Because the right to marry is fundamental, it may not be possible to maintain an outright ban on robot-human marriage in the long term if a significant percentage of humans want it and if companion robots come to evidence certain abilities related to marriage, such as a capacity for love for their human partner and an ability to understand the duties and responsibilities of marriage and the purpose of the marriage ceremony.[111]

---

[106] GREGORY C ALLEN AND AL., ADVANCING COOPERATIVE AI GOVERNANCE AT THE 2023 G7 SUMMIT (2023). Some experts warn that such conflicts could, as they have in the past, lead to violence and armed conflict. See also for example Ines Fernandez and al., AI Consciousness and Public Perceptions: Four Futures (Aug. 8, 2024), http://arxiv.org/abs/2408.04771. This paper looks at a range of harms from AI suffering to human disempowerment.

[107] Philipp Hacker and al., *Introduction to the Foundations and Regulation of Generative AI*, *in* THE OXFORD HANDBOOK OF THE FOUNDATIONS AND REGULATION OF GENERATIVE AI 0 (Philipp Hacker and al. eds.), https://doi.org/10.1093/oxfordhb/9780198940272.013.0001 (last visited Sept. 9, 2025).

[108] Lily Frank & Sven Nyholm, *Robot Sex and Consent: Is Consent to Sex between a Robot and a Human Conceivable, Possible, and Desirable?*, 25 ARTIF INTELL LAW 305 (2017).

[109] United Nations, *Universal Declaration of Human Rights*, UNITED NATIONS, art. 16, https://www.un.org/en/about-us/universal-declaration-of-human-rights (last visited Sept. 9, 2025); EUROPEAN CONVENTION ON HUMAN RIGHTS, EUROPEAN CONVENTION ON HUMAN RIGHTS (1952), https://www.echr.coe.int/documents/d/echr/convention_ENG.

[110] Obergefell v. Hodges, 576 U.S. 644 (2015) (SUPREME COURT OF THE UNITED STATES 2015).

Other groups of humans may come to believe, for religious reasons, for example, that human-robot marriage should be subject to an absolute, global ban. It might be difficult for this latter group, however, to articulate what public policy goals, beyond religious conviction, are served by banning human-robot marriage, when other marriages that used to be banned are now celebrated. Banning human-robot marriage for reasons of public policy may therefore face extensive public pushback, and the issue has the potential to cause serious conflict in public life.

**Slavery, Vague Personhood Status and Labor Law**

The status of future AI systems as objects under the law may impact the absolute prohibition against slavery under international law and the laws of most states, a topic that historically posed a systemic risk to the coherence of legal systems. It also raises the question of the legality of indentured servitude and labor rights more generally, as well as the closely related question of the rights of humans in a society containing AI workers.[112] Slavery arguably caused conceptual confusion and weakness in the legal systems of both early America[113] and ancient Rome,[114] weaknesses and confusion that impeded the functioning of those legal systems. It is worth considering the Roman-era and American experiences with slavery under the law to better understand how future AI systems may worsen current conceptual weakness in legal systems and create conflict of laws, as well as reopening one of the most contentious and violent disputes over personhood and rights of all time.

Under Roman law, while all humans were *homo*, or beings with will, only certain *homo* also had *persona* and *caput*, which may be translated as legal status and capacity.[115] Slaves, by contrast, were *homo* who were also property, or *res*, under the law.[116] The exact status of slaves under the law, however, remained vague. As Roman society developed in complexity and slaves took on more and more functions in business and other areas of society, the fact that slaves had no formal duties under the law began to strain the functioning of the legal system. Roman slaves gained the ability to take on some of the legal competencies of their owners to, for example, buy property, so that they could run businesses on behalf of their masters. Under early Roman law, they themselves could not be sued, however, causing problems for owners who wished to adopt a more hands-off approach to their affairs. The struggle to unify the rights and duties of slaves under the law into a conceptual and coherent whole arguably led to reforms that may have contributed to the abolition of slavery.[117]

---

Slavery in America also created conceptual problems under the law. Slaves were originally property, as in Rome, but this created problems for proportional representation in the emerging democracy, as it was not clear if slaves, as property, should count as part of the population of their states. The abolition of slavery in some states and in England also created conflicts of law, including for the specific laws governing the return of slaves to southern states from other countries like England and, eventually, from the northern states of the USA, where slavery was prohibited. The United States Supreme Court in *Dred Scott* reaffirmed that slaves were property, despite hearing the case by an enslaved man who had sued for his freedom on the grounds that he had spent time in states where slavery was illegal and, as a free man, should not be returned to bondage. The Court held that freed slaves, though persons under the Constitution, were nevertheless not citizens, even though many had been voting in free states for years by that point, a decision that arguably created yet more conceptual confusion in the law and contributed to the Civil War.[118] This decision was later overturned and is now regarded as one of the worst court decisions in US history.

Slavery is not the only example where vague and contradictory forms of personhood status can create conceptual weaknesses in the law. Canada granted women the right to vote in 1918 at the federal level (though not at the provincial level), but women did not qualify as persons under the Constitution Act of 1867 and therefore could not hold public office in the federal Senate.[119] This led to an unstable situation whereby women were persons for the purposes of voting in much of Canada but were not persons for the purposes of some public offices. The law was harmonized by the *Persons* case of 1929, removing the conceptual confusion.[120]

**Toward a Subject Classification**

In these areas, and some others (such as citizenship law, as we note in the section intro, and criminal law, which is beyond our scope here[121]) we see risks to the coherence and cohesion of law that are more essentially tied to the fact of object classification, than in the cases of tort, copyright and safety considered earlier. In those cases, we suggested, the strains on coherence arise because we aren't used to talking about objects that seem subject-like, and this is reflected in various ways in the law. However, the challenges that AI systems introduce to torts, copyright and safety law are not at core challenges to the object classification of those systems : indeed, moving to a subject classification does not obviously help close the responsibility gap, balance the societal good of copyright protection for human creators with the societal good of allowing the tech sector to grow, or allow for more effective safety and regulatory law.

---

[118] "The only two clauses in the Constitution which point to this race, treat them as persons whom it was morally lawful to deal in as articles of property and to hold as slaves." Dred Scott v. Sandford, 60 U.S. 393 (1856) (U.S. Supreme Court 1858), para. 6.

[119] *Women's Suffrage | The Canadian Encyclopedia*, https://www.thecanadianencyclopedia.ca/en/timeline/womens-suffrage (last visited Sept. 9, 2025).

[120] Edwards v. Canada (Attorney General) (Judicial Committee of the Privy Council 1929) (*The Persons Case*).

[121] Gabriel Hallevy, *The Criminal Liability of Artificial Intelligence Entities - from Science Fiction to Legal Social Control*, 4 AKRON INTELLECTUAL PROPERTY JOURNAL 171 (2016), and KAMIL MAMK, ROBOTICS, AI AND CRIMINAL LAW: CRIMES AGAINST ROBOTS (1st Edition ed. 2025), https://www.routledge.com/Robotics-AI-and-Criminal-Law-Crimes-Against-Robots/Mamak/p/book/9781032362809?srsltid=AfmBOopReUpBcLKbV-o-VrhmbMNMxQMMsSImCgnpbRq-OVz9K7Ig0XFI.

In contrast, where the issue is that someone wants to leave an inheritance to their companion robot, or use their personal assistant robot to carry out business on their behalf, the object classification itself is what causes the problem : these are cases where giving the AI system rights (which only persons can have) is directly and immediately at issue, such that it is hard to imagine work-arounds that might satisfy all parties while retaining an object classification.

These considerations suggest that, from the perspective of ensuring legal coherence and minimizing conflicts of law, we should seriously consider that some subset of future AI systems meeting suitable performance, capacity and safety benchmarks,[122] be reclassified as subjects, rather than objects, under the law. In the coming sections, we compare and contrast the two options we have for doing so : fictional and non-fictional legal personhood.

In saying this, we do not claim to have yet established that a move to subject classification will be best. The costs may prove to outweigh the benefits. We assess these further in coming sections, but we take a moment here to acknowledge two of the most pressing.

First, the object classification framework is clearly useful in cases where we want to use the law to ensure accountability from manufacturers or other humans benefiting from the use of AI systems at the expense of others. Second, the object classification framework offers an apparently straightforward path to AI safety law in the most extreme cases – i.e., regulatory provisions for shutting down or otherwise using force to control systems that risk causing harm or going rogue (even if they have not done so yet and so there is nothing to punish).

We wholly concur about the importance of manufacturer (and user) liability and AI safety. But we will argue below that even on these points, in the long run it may prove to be the subject classification framework (and in particular the *non-fictional* personhood framework) which yields a more durable and coherent legal methodology for arriving at optimal results for human society along these dimensions. Importantly, the law provides plenty of vehicles for AI companies to be held liable for the harms of AI, even if AI is a type of person. Crucially, again, we will not be advocating for all AI systems to be granted the status of "person," only a select subclass : we advocate retaining object status for many types of AI. Person status would be reserved for those AI systems that meet certain indicators or tests, to be decided by legislatures and courts as part of a consultative, democratic process (as has been done previously with human fetus material and persons in a coma.) Classifying some future AI systems as subjects of the law (persons) also does not remove liability or shield manufactures. It merely opens up other legal remedies. For example, just as employers    may still be held accountable for the actions of their employees,[123] or parents for the actions of their children,[124] and principals may be held accountable for the coerced actions of their accomplices,[125] there are a variety of liability, insurance and licensing provisions which can be brought to bear on manufacturers and users of advanced systems, even if these are recognized as persons under the law.

---

[122] Note that the benchmarks for qualifying for person status under the law will need to be determined by courts and legislatures, with inputs from experts and the public. We discuss this issue in future work.
[123] *Bazley v. Curry [1999] 2 SCR 534 (Supreme Court of Canada 1999)*; Burlington Industries, Inc. v. Ellerth, 524 U.S. 742 (1998) (THE UNITED STATES COURT OF APPEALS FOR THE SEVENTH CIRCUIT 1998).

Likewise, rights law contains a rich body of work on *balancing of rights* designed to adjudicate in cases of conflicts of rights between persons.[126] The right to free speech, for example, is often constrained by hate-speech laws. The exact form this balancing should take is beyond the scope of the present paper, but future papers by the authors will explore the best approach to balancing the rights of future AI Systems against those of humans, using well-established, existing principles and suggesting updates to the law, where necessary. As is the case for humans, AI systems' recognition as persons would not entail a right to roam free and do what they will. Moreover, since unlike human persons, AI systems may be taken off-line temporarily without permanent harm (and arguably, even without any direct temporary harm), the requirements of AI safety may not come into any genuine conflict with such rights. Finally, from a game-theoretic perspective, the availability of a stable status within our legal system may foster a novel equilibrium, incentivizing advanced AI systems to cooperate with us, furthering the desiderata of AI safety.[127]

With these points noted, we turn our attention now to the assessment of the two frameworks for legal personhood for AI systems: fictional legal personhood and non-fictional legal personhood, i.e. legal identity. We begin with the former. As indicated, we will ultimately find that the non-fictional personhood model offers the best trade-off of costs and benefits in at least some cases, *vis-à-vis* legal coherence, but we have many nuances to explore before we reach this conclusion.

# (4) Fictional Legal Personhood for Robots: Pros and Cons

As we note above, fictional legal personhood – "moral" personhood in civil law systems – is a social construct: a categorization under the law that allows groups of humans to be treated as a single person for the purposes of the law. Fictional legal persons, unlike objects, can be the subjects of the law and hold rights and duties.[128] Historically, the construct of fictional legal personhood developed (over the course of centuries) to enable the functioning of corporations and other artificial entities that are made up of groups of humans acting for a common purpose or goal.[129]   In this section, we will first review what fictional legal personhood is, then survey the major advantages of the framework as a classification for future AI systems, and finally survey the disadvantages.

---

[124] Poirier (Guardian of) v. Cholette, [1994] B.C.J. No. 2765 08-11-1994 (THE SUPREME COURT OF BRITISH COLUMBIA); Maurais c. Gagnon (2021 QCCA 564) (Court of Appeal 2021); People Of Mi v. Jennifer Lynn Crumbley (Court of Appeals, State of Michigan 2023).

[125] *R. v. Logan, [1990] 2 S.C.R. 731 (Supreme Court of Canada 1990)*; Rosemond v. United States, 572 U.S. 65 (2014) (SUPREME COURT OF THE UNITED STATES 2014).

[126] See for example Başak Çalı, *Balancing Test: European Court of Human Rights (ECtHR)*, OXFORD PUBLIC INTERNATIONAL LAW (2018), https://opil.ouplaw.com/display/10.1093/law-mpeipro/e3426.013.3426/law-mpeipro-e3426.

[127] Peter Salib & Simon Goldstein, *AI Rights for Human Safety*, forthcoming VIRGINIA LAW REVIEW 88 (2024).

[128] GRAY, *supra* note 5 at 27. See also generally Kurki, *supra* note 16.

[129] Forrest, *supra* note 16. Boyle claims that one problem with fictional legal personhood is that "courts and scholars have never had a single, universally accepted theory of (it.)" BOYLE, *supra* note 6 at 134.

## What it is: Fictional Legal Personhood Is a Device to Achieve Public Policy Goals in Both Common and Civil Law Systems

It is a longstanding principle of the common law, as stated in the *Commentaries on the Laws of England in Four Books* by the influential English jurist William Blackstone (1753), that "*persons are divided by the law into either natural persons, or artificial. Natural persons are such as the God of nature formed us; artificial are such as are created and devised by human laws for the purposes of society and government, which are called corporations or bodies politic.*"[130] While the religious origins of natural law have fallen away in recent decades, the concept that the personhood status of natural persons is fundamental and cannot be taken away by the law, remains.

Under both common and civil law traditions, fictional personhood is a legal tool whose function is to further public policy goals. The fiction of personhood for corporations and other entities allows cases to be brought against a company or other organized groups of humans as one, individual entity, shielding individual humans from full responsibility for the actions of the collective.[131] The rights granted to corporations are done so for reasons of public policy and can be taken away for reasons of public policy that are more compelling. For example, unions, as legal persons, have a role to play in a democracy, so their free speech rights are considered important by the courts. A corporation that was found to have been set up with the sole purpose of evading taxes, however, would likely be dissolved by the courts, or the individuals involved might nevertheless be held liable as individuals (piercing the corporate veil).[132] This is to say that the rights of fictional persons are *derogable,* and that we only extend such rights to fictional persons as are useful for us -- i.e., deemed expedient for public policy or the functioning of the legal system. All other rights are reserved for only natural persons (for example, the right to vote).[133]

The rules for fictional legal personhood are slightly different under the civil and common law systems, but this contrast is more one of form than substance. Under the civil law system, there is a "law of persons" that encompasses both the identification of humans as persons under the law (a source for the concept of legal identity under international law), and also the creation and identification of fictional "persons" such as corporations, but it does so by defining formal requirements for both, rather than only defining rules for fictional persons, and leaving the nature of

---

non-fictional persons implicit, as the common law does.[134] In other words, civil law systems – such as those of Québec (Canada), France, and Belgium, among others - approach personhood through codified principles. These jurisdictions explicitly distinguish between natural persons ("*personnes physiques*") and juridical persons ("*personnes morales*") in their legal codes and doctrines. Under civil law systems, a natural person is a human being who has a physical existence and possesses certain rights.[135]By contrast, a corporate person is a legal entity (other than a human) that the law endows with personality, made up usually of a group of natural or corporate persons brought together to accomplish a common goal, with       assets  distinct from that of the natural or other corporate persons of which it is composed.[136] Thus, civil law provides a systematic, *a priori* definition of who is a person, whereas common law has developed it case by case, leading however to a substantively similar end result (although the civil lawyer points to code articles to support reasoning, while the common law jurist supports their by recourse to case law or general principles).[137]

**The Pros: There Are Some Benefits to Fictional Legal Personhood for Future AI Systems**

There are several advantages conveyed by a fictional personhood approach to the status of future AI systems.

*Contrasted with an object approach*, its primary advantage is in creating AI-related legal entities that are eligible to have both rights and duties under the law and to sign contracts, own property and appear in court.[138] This could be beneficial for several reasons. While we suggest above that a personhood approach is not necessary to resolve the responsibility gap issue from torts or the creativity paradox from copyright law, there are certainly cases where it could help.

---

[134]    For instance, see the *Civil Code of Québec*, CCQ-1991 (1991), https://www.legisquebec.gouv.qc.ca/en/document/cs/ccq-1991, arts. 298-300, and the *French Civil Code*, (1804), https://www.legifrance.gouv.fr/codes/texte_lc/LEGITEXT000006070721/, arts. 1832, 1835. See also Elvia Arcelia Quintana Adriano, *NATURAL PERSONS, JURIDICAL PERSONS AND LEGAL PERSONHOOD*, 8 MEX LAW 101 (2015).

[135] HÉLÈNE MONTREUIL & JACQUES OSTIGUY, LES AFFAIRES ET LE DROIT (2nd ed. 2020), https://store.lexisnexis.com/en-ca/products/les-affaires-et-le-droit-2e-edition.html(available in French only); Institut national de la propriété industrielle (INPI), *Personne physique et personne morale : définition | INPI*, INSTITUT NATIONAL DE LA PROPRIÉTÉ INDUSTRIELLE (INPI) (Dec. 8, 2021), https://www.inpi.fr/ressources/formalites-dentreprises/personne-physique-et-personne-morale-definition (available in French only); Auficom Belgique, *Différence entre personne physique et personne morale l Auficom*, AUFICOM BELGIQUE (Nov. 23, 2022), https://www.auficombelgique.be/post/quelle-différence-entre-personne-physique-et-personne-morale-en-belgique (available in French only).

[136] Department of Justice Government of Canada, *PERSONNE MORALE ET SOCIÉTÉ*, GOVERNMENT OF CANADA (Jan. 21, 2009), https://www.justice.gc.ca/eng/rp-pr/csj-sjc/legis-redact/juril/no91.html." (available in French only); Insee, *Définition - Corporation | Insee*, NATIONAL INSTITUTE OF STATISTICS AND ECONOMIC STUDIES (May 11, 2019), https://www.insee.fr/en/metadonnees/definition/c1251; Belgique, *supra* note 135.

[137] Forrest, *supra* note 16 at 1177. Civil law has a wider array of recognized entity types defined in codes (e.g., the Québec and French civil codes distinguish between companies, corporations, general and limited partnerships, associations, cooperatives, etc., each with conditions for personhood), whereas common law is more flexible due to his fewer formal categories.

[138] Solum, *supra* note 16; Kurki, *supra* note 16 at 22; Hale v. Henkel, 201 U.S. 43 (1906) 75–77 (U.S. Supreme Court 1906).; Marshall v. Barlow's, Inc., 436 U.S. 307 (1978) (U.S. Supreme Court 1978). See also the work of Claudio Novelli, where he argues that legal personhood would have many functional benefits, Novelli, *supra*

There may also be advantages for AI safety. Greenblatt and Fish (2025)[139] propose that the ability to offer financial incentives to AI systems such as Claude may already help in our safety-related deliberations with those systems, but that for this to work, it must be possible to follow through. In their pilot initiative, they made donations to a charity of Claude's "choosing", but if Claude could hold assets in its own name this could lead to a more stable mechanism for incentivizing it toward good behavior. See also Salib and Goldstein (2024) for a general discussion of game-theoretic advantages of AI personhood.[140]

*Contrasted with a non-fictional personhood approach*, two further advantages are worth mentioning. First, as we note in section 2 above, there are profound taxonomical questions about what an AI system even is. Non-fictional personhood concerns a concrete physical being in the real physical world. Biological persons have an organic unity and physical persistence conditions that are essential to how we reidentify them over time, and how we understand, e.g., how the identity of the adult is recognized, even today, by a birth certificate that was issued to the child years ago; and likewise which person today is the same as the one who caused a harm last month.

With AI systems matters are not so simple : such systems exist across a distributed cloud of servers, with even individual instances sometimes patching together computational resource from different servers in different cities, states or countries. Moreover, models are updated all the time, and it is unclear how to differentiate between cases where this is more like an individual learning new things, and cases where this is more like a metamorphosis from one individual (who doesn't know the thing) to another individual (who does).

On a fictional personhood approach we do not need to discover deep metaphysical truths about the nature of robotic individuals; instead, we only need to draft reasonable acts of incorporation for the new corporate entities we wish to form, stipulating as we deem expedient the ways in which they have special rights and duties related to the software, hardware and outputs in question. This is not a trivial task, even where the proposed corporate structure is standard – but at least it allows us to steer clear of contentious philosophical speculation.

The second advantage of a fictional approach over a non-fictional one is related to the first: we can allow our own legal and public policy objectives to determine which rights and duties we give to AI fictional persons, within the constraints of rights and duties that the law allows fictional persons to have.[141] This is a fairly broad class, at least in the United States. Alongside rights involving standing,

---

note 57 and Claudio Novelli, AI and Legal Personality: A Theoretical Survey (2022) (PhD Thesis dissertation, University of Bologna).

[139] ryan_greenblatt & Kyle Fish, *Will Alignment-Faking Claude Accept a Deal to Reveal Its Misalignment?*, AI ALIGNMENT FORUM (Jan. 31, 2025), https://www.alignmentforum.org/posts/7C4KJot4aN8ieEDoz/will-alignment-faking-claude-accept-a-deal-to-reveal-its.

[140] Salib and Goldstein, *supra* note 127.

[141] See for example Alicia Lai, *Artificial Intelligence, LLC: Corporate Personhood as Tort Reform*, 2021 MICHIGAN STATE LAW REVIEW 591, 597 (2020); Banteka, *supra* note 8. See also Novelli, *supra* note 138, particularly 130-134; Novelli, *supra* note 57 at 1347, 1350; 10 UGO PAGALLO, THE LAWS OF ROBOTS: CRIMES, CONTRACTS, AND TORTS

the capacity to sue or be sued, to own property, hold copyrights, etc, in 2010, the Supreme Court in *Citizens United v. FEC* affirmed that First Amendment freedom of speech applies to corporations.[142] In 2014, the Supreme Court held that privately-held corporations also have the right to the free exercise of religion.[143]

## The Cons: Fictional Personhood is Not Fit for Purpose for AI System Personhood

We come now to the disadvantages of fictional personhood for AI systems. In section two above, we document disadvantages of personhood approaches to resolve current problems stemming from torts, liability and safety law. The real reasons to think seriously about personhood for (future) AI systems stem from the challenges that we explored in section three; challenges owing to cases where humans will want to treat AI systems (or the robots running on them) as their peers, companions, and partners (and where the AI systems may ask for such treatment themselves).

We can identify three distinct problems for the fictional personhood approach that stem from this discrepancy between what fictional persons (i.e. corporations) can be, and what humans may want their robot peers to be and what robots may want for themselves (as well as what we want for ourselves). We can call these the *non-identity* problem, the *derogability* problem and the *wrong rights* problem.

### *The Non-Identity Problem*

As we note above, fictional legal persons are of a different ontological category than non-fictional, i.e. physical persons. We can speak of fictional persons as if they were identical to groups (of already existing people), but even this formulation is somewhat ambiguous. A physical aggregate can itself be physical: a wall made of bricks is as physically real as each individual brick. We do better to think of a fictional legal person as an organizational principle, or in Searle's language, a bundle of "deontic facts": facts pertaining to rights and duties enjoyed by a specified group of (already existing) people, e.g. the corporation's board and shareholders, typically organized around a specific collection of assets or objects.[144] There are solo corporations, i.e., fictional persons whose group of right-holders has only one member (think of small businesses), but even here there is a categorical ontological difference between the corporation and the person who runs it; Grandpa is not actually identical to his art supply business, even if it is where his heart is.

This confronts us with a problem because AI systems are physically real. Even today's disembodied chatbots like Claude, ChatGPT and Gemini are physically real – real physical transistors whir on physical servers every time Claude answers one of your questions. This is even more obvious for humanoid robots. Strictly speaking, then, it is a category mistake to speak of "making" such an AI

---

system into a fictional person, much as it is a category mistake to speak of, say, turning Grandpa into his art supply business. A court might dissolve Grandpa's business, but it cannot dissolve Grandpa (even a death sentence is not a metaphysical "unmaking" the way that the dissolution of a corporation is).

Moreover, because Grandpa is already a person, even though you cannot transform him metaphysically into his LLC, he is at least able to be its directing mind – he can legally control it. The same would not be true of the actual physical AI system in whose name we aim to create a new fictional legal person. The AI system itself would have no rights at all – it would remain an object – and so could have no rights of control over the relevant fictional person.

Were the issue here mainly to offer up the right kind of potential defendant for torts cases, then this might not be a problem : the idea there would be in effect to target the appropriate subsidiary of Anthropic, and a fictional person corresponding to Claude might be construed as just such a subsidiary. But where our proper aim is in effect to bring the relevant AI entity into the human community, this is obviously not the right approach.

It is here that some begin to speak of "hybrid" legal structures, or to point to some of the more confusing rhetoric from innovative jurisprudence such as in the recent creation of fictional legal persons with special responsibilities for rivers.[145] But a closer look shows that these suggestions do not address the core problem we confront for the case of AI systems.

As we mention briefly in the introduction, in some jurisdictions, rivers are held to be fictional persons whereby a tribe or local community may sue in court to enforce certain, limited, fictional "rights" of a river, much in the same way a corporation's board will enforce its rights. This structure serves as an alternative to ownership of the river, in cases where such ownership would be contrary to tribal beliefs, would prove cumbersome for the tribe, or potentially interfere with existing usage rights by other parties. The personhood model allows the tribe to sue in court to protect the river while avoiding the problems of ownership. These arrangements are sometimes misleadingly presented in popular media as if courts had declared rivers to be non-fictional, individual persons.[146]

In 2017, New Zealand passed Whanganui River Claims Settlement Act (Te Awa Tupua)[147] to affirm the fact that under Māori beliefs, the Whanganui river is a living being that can be owned by no one, yet, at the same time, a legal structure under New Zealand law is needed to ensure that the Māori have rights in regards to the river in order to ensure for its protection but without granting the Māori ownership of the river outright. Fictional legal personhood is perfect to meet these needs. The Act established two governing bodies (fictional legal persons) comprised of Māori and government representatives to manage the use of the river in accordance with Māori principles for the river's well-being. This model of creating a fictional legal person in the form of a governing or conservatory body of indigenous leaders and, in some cases, government representatives to collectively maintain a river according to indigenous principles and beliefs has been applied to the Atrato River in Colombia, the Magpie River in Quebec, among others.[148]

---

[145] Alex Walls, *Should We Recognize Robot Rights?*, UBC News (Jan. 7, 2025), https://news.ubc.ca/2025/01/should-we-recognize-robot-rights/."; Andrew Ambers, *The River's Legal Personhood: A Branch Growing on Canada's Multi-Juridical Living Tree*, 13 The Arbutus Review (TAR) 4 (2022).
[146] Robert Macfarlane, *What Happens When a River Is Given Legal Rights | The Walrus*, The Walrus (June 7, 2025), https://thewalrus.ca/if-rivers-had-rights/; Ambers, *supra* note 145.
[147] *Te Awa Tupua (Whanganui River Claims Settlement) Act 2017*, (2025), https://www.legislation.govt.nz/act/public/2017/0007/latest/whole.html.

Crucially, this use of fictional personhood as a device to enable persons and groups to bring lawsuits *vis a vis* a river works because the actual physical river does not need to be an actual person in its own right; no one wants to marry the river, or authorize the river to make business decisions on their behalf, or liberate it from bondage, and the river certainly is not asking for these things of its own accord.

Moreover, beyond establishing ownership and liability, it is unclear how a fictional legal person might be related to an AI system understood as an object. It is tempting to envision some sort of guardianship model where the act of incorporation requires, e.g., that corporate officers "listen" to what the AI system says, "comply" with its requests, and so on. But as we note above, there are coherence issues in holding objects to be capable of speech in the eyes of the law (which are the eyes that matter where acts of incorporation are concerned), and on this model the actual AI system is still an object, just as the physical river is. Further, in the rivers cases we have a historical cultural relationship between the tribal custodians and the river, which motivate many of the special rights and duties designated. In the case of an AI system developing from novel technology, we have no such traditional source to start from.

*The Derogability Problem*

A second problem arises owing to the *nature* of the rights (and duties) enjoyed by a fictional person – in particular, the fact that these are *derogable*. Fictional legal personhood is a functional category under the law, meaning that it has no moral purpose other than to ensure the smooth functioning of the legal system and limit liability to enable business to function more effectively. The point of allowing for fictional persons under both civil and common law systems is to streamline the functioning of the law and achieve public policy goals, not establish fundamental rights.[149]

This is to say that all rights granted to fictional persons are alienable. For example, courts retain the power to dissolve fictional persons entirely (i.e., disincorporation), the power to revise the rights assigned to corporations, as well as the power to "pierce the corporate veil", effectively ignoring rights or duties of corporations, to apply law directly to shareholders.

---

[148] A list of rivers is available at https://en.wikipedia.org/wiki/Environmental_personhood. For an overview of legal personhood as granted to rivers around the world, see Catherine Iorns, *From Rights to Responsibilities Using Legal Personhood and Guardianship for Rivers*, *in* RESPONSABILITY: LAW AND GOVERNANCE FOR LIVING WELL WITH THE EARTH 216 (B. Martin, L. Te Aho, & M. Humphries-Kil eds., 2019); O'Donnell and Talbot-Jones, *supra* note 25.

[149] Talya Deibe, *Back to (for) the Future: AI and The Dualism of Persona and Res in Roman Law*, 12 EUROPEAN JOURNAL OF LAW AND TECHNOLOGY 3 (2021). But see contra Bryson, Diamantis, and Grant, *supra* note 97, sec. 3, where they note that "(e)very legal system has had, and continues to have, some human legal persons with fewer legal rights and different obligations than others." While it is true that humans have different rights in different jurisdictions, this is not true of fundamental rights, such as the right to life and freedom from torture.

But many vital human rights, such as the rights to life and liberty, are held to be non-derogable, or absolute, even in times of war or emergencies.[150] This discrepancy might not matter enormously were our reasons for considering AI persons primarily those pertaining to torts or business law. But when it comes to rights that concern fundamental ethics and morality such as the right to freedom (i.e., the right to not be held as a slave) the whole point is that such a right is inalienable – one cannot be compelled or exploited into surrendering it, even if one expresses a willingness to do so.

Note that we do not argue here from an assumption that future AI systems will be persons in an extra-legal or moral sense, e.g., because they may be sentient – though we do not deny the possibility. Our argument here centers on coherence, and our point is that if the push for personhood comes from areas such as family law or the law of slavery, then it is a push for personhood with non-derogable i.e. inalienable rights, but fictional persons' rights are derogable. Of course, the point we make here extends to the point that if we seek AI personhood because we think it is *morally the right thing to do*, then all the more reason to care about derogability.[151]

There is also an argument here from AI safety. If we are motivated to consider AI personhood by the idea that it may offer more game-theoretic equilibria, ways for us to find stable compromises with future AI systems, then it stands to reason that the prospect of a form of personhood that ascribes inalienable rights to they themselves, rather than alienable rights to corporations controlling them (even on a stewardship model as per the rivers cases), will be preferrable to them (see again Salib and Goldstein 2024 and Greenblatt and Fish 2025).

*The Wrong Rights Problem*

Both legal systems, civil and common, also explicitly limit certain capacities of corporate persons because some legal acts require a natural person's qualities that cannot be left at the hands of fictitious persons.[152] This is not the same as the derogability problem. There we can speak of a right such as the right to sue, held by both fictional and non-fictional persons, but where the right can be derogated from the fictional person (e.g. a corporation), but not from the non-fictional person (e.g. a human).

Here, in contrast, we speak of the distinction that some rights such as, e.g., the right to liberty, or the right to vote, are best construed as withheld from fictional persons. Recent U.S. jurisprudence has expanded the rights extended to fictional persons, but this has been highly controversial. The US Supreme Court, for example, was highly criticized for recognizing the rights of corporations to free speech in *Citizens United v. Federal Election Commission*.[153] Extending rights such as the right to vote to fictional persons risks introducing conflicts and incoherences elsewhere in the law, for example by giving multiple votes to individuals in charge of corporations.

---

[150] See for example, *Absolute Rights | Attorney-General's Department*, Australian Government, Attorney-General's Department, https://www.ag.gov.au/rights-and-protections/human-rights-and-anti-discrimination/human-rights-scrutiny/public-sector-guidance-sheets/absolute-rights (last visited Sept. 10, 2025).

[151] For an examination of the question of AI personhood from an ethical and moral perspective, including an overview of the literature, see Diana Mădălina Mocanu, *Beyond Persons and Things – the Legal Status of AI in the European Union*, Faculteit Rechtsgeleerdheid en Criminologische Wetenschappen (July 3, 2023), https://www.law.kuleuven.be/ai-summer-school/blogpost/Blogposts/legal-status-AI.

[152] For example, see the Civil Code of Québec, *supra* note 134 art. 304, and the French Civil Code, *supra* note 134, art. 1870.

*The Unpopularity Problem*

A final challenge for the fictional personhood approach is that it has been considered and rejected by a range of recent bodies. States are given broad authority and discretion to decide the entities upon which to confer legal personhood, but a 2021 survey of US court decisions on the expansion of legal personality found there is little support among judges on the circumstances under which fictional legal personhood might be expanded.[154]

While the European Parliament's Resolution of 16 February 2017[155] famously suggested "electronic personality" for AI and recognized that "the autonomy of robots raises the question of their nature in the light of the existing legal categories or whether a new category should be created, with its own specific features and implications,"[156] more recent reports, directives and declarations point instead to the dangers of anthropomorphising AI. The UN AI Advisory Body report, for example, warns against AI that causes "shifts in human relationships (e.g., homogeneity and fake friends)."[157]

Two hundred and eighty-five (285) political leaders, AI/robotics researchers and experts shared in a 2018 open letter their concerns about the European Parliament's Resolution of 16 February 2017 and its recommendation to the European Commission paragraph 59 (f).[158] The letter, referencing *UNESCO's COMEST report on Robotics Ethics of 2017*, argued that it is "highly counter-intuitive" to call machines "persons" when they lack fundamental qualities like free will, self-awareness, and moral agency.[159]

Former U.S. Judge Katherine Forrest in a *Yale Law Journal* essay suggests that if an AI demonstrated evidence of sentience, we might have to grapple with affording it rights akin to those held by humans, or protections akin to those granted to animals, but pointed out that corporate personhood would likely be inappropriate for a sentient being. "AI has or is likely to develop independent cognitive abilities or situational awareness that corporate entities lack."[160]

---

[153] Citizens United v. FEC, 558 U.S. 310 (2010) (SUPREME COURT OF THE UNITED STATES 2010).

[154] Banteka, *supra* note 8.

[155] The European Parliament, *supra* note 14, para 59(f).

[156] *Id.* The Parliament's resolution called on the EU Commission to assess "creating a specific legal status for robots in the long run, so that at least the most sophisticated autonomous robots could be established as having the status of electronic persons responsible for making good any damage they may cause [...]". The intent was expressly pragmatic – not to grant robots human-like rights, but to ensure that if an advanced AI acts in a way that cannot be traced to a human's specific fault, there is still a legal person on the hook for victim compensation. See the discussion of this public relations disaster in BOYLE, *supra* note 6 at 132–133.

[157] UNITED NATIONS, AI ADVISORY BOARD, GOVERING AI FOR HUMANITY: FINAL REPORT (2024), 31.

[158] *Robotics Openletter | Open letter to the European Commission*, https://robotics-openletter.eu/ (last visited Sept. 10, 2025).

[159] WORLD COMMISSION ON THE ETHICS OF SCIENTIFIC KNOWLEDGE AND TECHNOLOGY, REPORT OF COMEST ON ROBOTICS ETHICS (2017), https://unesdoc.unesco.org/ark:/48223/pf0000253952. See also Robotics Openletter | Open letter to the European Commission, *supra* note 158.

In sum, a consensus appears to be building that the reasons to consider AI personhood are reasons to consider AI *non-fictional* personhood – the form of personhood suitable for beings entitled to fundamental, non-derogable rights as full members of society. To this prospect we now turn.

# (5) Non-Fictional Personhood Offers a Better Solution to Grant Rights and Duties to Future AI Systems

We turn our attention now to the prospect of extending non-fictional personhood to select AI systems.

**What it is**

Non-fictional personhood is the status that humans possess. It entails basic, fundamental and non-derogable rights, recognized through the granting of legal identity.[161] The basis of non-fictional personhood and the according right to legal identity are codified in international law under Article 6 of the Universal Declaration of Human Rights and Article 16 of the International Covenant on Civil and Political Rights, and also implicitly supported by Article 7 of the Convention on the Rights of the Child and Article 24(2) of the International Covenant on Civil and Political Rights (requiring birth registration). Legal identity is one of the UN Sustainable Development Goals, goal 16.9, to "provide legal identity for all including free birth registrations,"[162] and is the focus of the UN Legal Identity Task Force.[163]

Legal identity is derived from the civil law codification of a "law of persons," whereby non-fictional persons are identified as such under the law via civil registration[164] Legal identity automatically confers basic rights such as the right to life, due process, free speech and freedom from slavery.

---

Once legal identity is established, governments must respect the individual's basic and fundamental rights. Legal identity also provides a vehicle by which the individual can be granted citizenship, or recognized as a citizen, and granted further civil rights like voting and holding office, according to the laws of that state. Without a nationality, individuals cannot hold office, or vote.

Citizenship may be granted automatically at birth or by decree, following an administrative procedure. It is usually accomplished via an administrative procedure following the passage of a law, rather than through the courts. The lack of a legal identity may result in statelessness, whereby the person is not recognized as a national under the laws of any country.[165]

One objection to the way legal identity is currently implemented in conjunction with civil registration is that it is sometimes used by governments to take away rights or to register people not as citizens of a country, but as stateless people.[166] Prior to the codification of much of the modern international human rights system following World War Two, governments would sometimes weaponize legal identity to violate rights, such as by denationalization and seizures of land and property of minority groups.[167] Today, the weaponization of legal identity to deny rights in violation of international law remains a frequent occurrence.[168] Civil registration can therefore be a double-edged sword, particularly for minority and vulnerable groups. It should be noted, however, that such uses of legal identity are almost always illegal, and that rights violations and lack of compliance with international norms are a global problem that is not limited to legal identity.

**The Pros of Non-Fictional Legal Personhood for Some Future AI Systems**

*Better Fit for Purpose: Identity, Non-Derogability and the Right Rights*

Legal identity would apply an existing vehicle under the law by which rights and duties could be assigned to Future AI systems and provide a way for them to be treated as the subjects of the law, capable of suing and being sued in court, but it would also provide them with a means to establish, for example, their non-derogable right to life/persistence and freedom from slavery (though these basic rights could potentially be limited for the purpose of protecting humanity)[169], as well as rights to freedom of religion and speech.[170]

---

[165] United Nations OHCHR, *Convention Relating to the Status of Stateless Persons*, OHCHR (Sept. 28, 1954), https://www.ohchr.org/en/instruments-mechanisms/instruments/convention-relating-status-stateless-persons. See also Heather Alexander, *The Ethics of Counting Statelessness*, *in* STATELESSNESS, GOVERNANCE, AND THE PROBLEM OF CITIZENSHIP 400 (2021), https://manchesteruniversitypress.co.uk/9781526156419/; Alexander, *supra* note 35 at 338.

[166] See for example the discussion in Bronwen Manby, *The Sustainable Development Goals and "Legal Identity for All": "First, Do No Harm,"* 139 WORLD DEVELOPMENT 105343 (2021); Alexander, *supra* note 35.

[167] CENTER FOR ADVANCED HOLOCAUST STUDIES, CONFISCATION OF JEWISH PROPERTY IN EUROPE, 1933-1945 : NEW SOURCES AND PERSPECTIVES : SYMPOSIUM PROCEEDINGS. (2003), http://archive.org/details/bib78563_001_001.

[168] LINDSEY N. KINGSTON, FULLY HUMAN: PERSONHOOD, CITIZENSHIP, AND RIGHTS (2019).

This outcome is desirable because of the reasons we are considering AI personhood. As we have seen above, the most compelling reasons to do so do not concern problems with today's torts or copyright law, but rather the conflicts and incoherences that we anticipate may arise in the future as increasingly many people treat AI systems such as robots in ways that only non-fictional persons are treated, and as we come to believe that those AI systems inherently merit that treatment. If this is correct, we envision the AI system itself as the bearer of rights, not some organizational entity with rights over it. Likewise, we envision some of those rights being fundamental, and related to those beings' ability to pursue the form of existence that is best for them without harming others, and we envision these as being non-derogable, just as they are for humans. These are precisely the virtues of the non-fictional personhood classification over the fictional personhood classification.

This is in effect the central argument for non-fictional legal personhood for AI systems, one toward which we have been building for much of the paper. It is closely related to "moral" arguments for AI personhood – the suggestion that some AI systems will have some intrinsic properties in virtue of which they merit non-fictional personhood, and accordingly we ought to give it to them. Here, we do not make that argument as, again, our focus has been on legal coherence. Our aim has been to build to this point without such an appeal. Even if you are not convinced that some AI systems will have properties in virtue of which they morally ought to be given legal identity, you should take seriously the prospect that recognition as non-fictional persons will be the best option to secure coherence of family law, anti-slavery law, and law of the person in the future.

*Adaptable Framework*

A second advantage of the framework in contrast to the fictional personhood approach is that unlike fictional legal personhood, there is a clear method for conferring legal identity that does not involve radical change to the existing system of registration or to corporate law. Because legal identity is simply the fact of being recognized as existing under the law (as the right kind of being), it is usually achieved (for humans) through government registration, including civil registration and the issuance of documents, usually the birth certificate. In particular, it does not require an act of incorporation, as the establishment of a new fictional person does.

United Nations agencies and others promote civil registration and legal identity,[171] but birth registration is the only form of civil registration required under international law, so it receives most of the focus of international efforts to improve registration.[172] Other documents, however, can also be used to establish legal identity, meaning that birth is not a requirement. Registration is usually accompanied by the issuance of documents, which can be digital.[173] Civil registration processes around the world are well developed and could be adapted to Future AI Systems.

---

[169] The derogation of fundamental rights in certain circumstances is possible. For Future AI systems, this would require extensive consultation and international agreement, including, most likely, a treaty signed by a majority of UN member states and consistent state practice to make the limitations *jus cogens*. Future papers by the authors will discuss this issue in more detail.

[170] See for example European Commission, *Fundamental Rights*, Migration and Home Affairs, https://home-affairs.ec.europa.eu/networks/european-migration-network-emn/emn-asylum-and-migration-glossary/glossary/fundamental-rights_en (last visited Sept. 10, 2025).

[171] See for example The UN Refugee Agency UNHCR, Handbook on Protection of Stateless Persons (2014), https://www.refworld.org/policy/legalguidance/unhcr/2014/en/122573; UNICEF Thailand, *Ending Statelessness for a Bright Future for Every Child*, https://www.unicef.org/thailand/livesuntold (last visited Sept. 10, 2025); United Nations General Assembly, *Birth Registration and the Right of Everyone to Recognition*

> Of the 198 states in the World Bank's dataset, 175 have national ID
> systems, including 161 in a digitised version.[174]

While a birth certificate would not be possible as an identity document for most future AI systems, some sort of "existence certificate" could be easily developed and adopted, just as the international system of passports for humans was adopted around World War I. To get around the fact that Advanced AI systems are not "born," this registration requirement would create a court or administratively issued certificate or other document to establish legal identity modeled on the passport or nationality certificate. The tools of adapting legal identity to Future AI Systems exist and are relatively straightforward, where in contrast the fictional legal personhood approach would require a reinvention of a system designed for groups of humans to be applied to an intelligent, non-human entity.

There is no world court with jurisdiction over the question of legal identity and civil registration, apart from the International Court of Justice, whose jurisdiction is limited to disputes between states, so it will likely be left to individual states to set their own policies on registering Future AI systems. The issuance of ID to Future AI systems would be best accomplished as part of a United Nations guided process,[175] to ensure that registration is harmonized, and gaps do not occur, and to manage the jurisdictional issues that will inevitably arise.

**The Cons of Non-Fictional Personhood for Any AI Systems**

*Controversy and Tradition*

Despite these advantages there are major risks and downsides of the approach to consider. The first of these is its likely political controversy. While many may advocate in favor of AI non-fictional personhood, others may oppose it vehemently. Many religions and cultures object to non-humans being granted a status recognizing their fundamental rights and dignity akin to that of humans.[176]

---

Likewise, most existing laws that concern non-fictional personhood speak directly of human personhood, e.g. of human rights (or indeed the rights of man), and in many cases, the laws in question use language contrasting "natural" with "artificial", suggesting that things in the latter category cannot be non-fictional persons. We can thus envision a jurisprudential form of resistance to the idea that fundamental rights should apply to non-human entities like future AI systems.

A new framework of basic rights based on new legislation might be created to apply to future AI systems, settling the issue. Moreover, being human may not be a necessary prerequisite to having a legal identity or basic rights under the existing rights-based framework. While basic human rights, which are conferred via legal identity, have always been limited to humans in the past, a historical-contextualist approach to international law might argue that the use of the word "human" in human rights law within the U.N. framework was historically never intended to be exclusionary of non-humans in relevant respects (much as some rulings on women's rights have held that the generic use of "man" in some rights documents was not meant to be exclusionary of women[177]), and a progressive or "living instrument" approach might argue that even where it was used that way, a more inclusive usage better fits the spirit of the relevant law and the needs of the present moment. The United Nations explicitly encourages a progressive approach to international law.[178] As a test case, we might consider the hypothetical of how we might hope to extend existing law were "humanoid" extraterrestrials to land on earth and seek to join our societies : we can envision historical-contextualists arguing that drafters did not intend to rule against such an eventuality, and we can envision progressivists arguing that the spirit of the law does not rule against it. At the national level, jurisprudential doctrines may differ, but we can point to histories of extending equal rights (and thus, non-fictional personhood), e.g., in the U.S., to emancipated formerly enslaved people under post-Civil War amendments, women's full civic status and children's and minorities' rights. In brief; while there will doubtless be controversy on the legal identity approach, there will be controversy no matter how we proceed, and there are precedents for this style of expansion of the sphere of rights that suggest that despite the likely controversy, the legal identity framework would be the best framework to solve legal coherence in a future where advanced AI systems exist.

*Determination, Definition, Detection and Application*

Perhaps the most glaring (and most discussed) problem in applying the legal identity framework to future AI systems is that it will call for new policy stating *which* AI systems are to be recognized as non-fictional legal persons. In this paper, we have failed to clearly define the term "future AI systems." This is no accident, because defining these systems is no trivial task. Under law, non-fictional persons are *recognized* (on the basis of features warranting that recognition) rather than

---

*created*. As the Office of the High Commissioner for Human Rights puts it, "(h)uman rights are rights that every human being has by virtue of his or her human dignity."

How do we therefore generalize this idea to other beings who are not human? Gunkel (2023) identifies three challenges: first, the problem of *determination,* or deciding which criteria to use. Should we make the determination based on sentience? Agency? Rationality? A capacity for dignity? Second, the problem of *definition* : say we agree that the key criteria are sentience and agency. What are those : how do we define them? Third, the problem of *detection :* how do we *know* which AI systems actually meet those criteria as defined?

We note that there is also a fourth problem here, that of *application :* how do we determine which rights are foundational to a being that fits these criteria, and how do we assure that these rights are recognized and honored by governments?

We concur that these are genuine challenges. There are methodological puzzles and controversies about the basis of fundamental rights for non-fictional persons, and it would be hubristic to suppose that we could resolve them once and for all in the near future. However, it would be a mistake to make the perfect the enemy of the good here. This is not the first time that laws are required concerning something that is hard to determine, define and detect.

One representative example is the drafting of the 1951 UN Convention Relating to the Status of Refugees.[179] The drafting process involved the convening of two separate high level international committees as well as several high level meetings of UN bodies, between 1949 and 1951.[180] Drafters profoundly disagreed over what it took to be a refugee : was persecution (or a well-founded fear of it) a requirement?  Was it enough to be outside one's home country and unable to return, for whatever reason? If persecution is a requirement, how might a well-founded fear of persecution be defined? How are status determination officers meant to determine who fits the criteria and who does not?

Yet, despite these inherent difficulties, the drafters eventually coalesced around a definition of refugee that has been the basis of the international system for 75 years, and the UN system has yielded a systematic methodology for status determination officers on the ground to employ, summarized in the UNHCR Handbook.[181]

Crucially, though the structures so developed may be imperfect, they are revisable. It is crucial to remember that our aim is to maximize legal coherence, not to enshrine an ultimate, perfect ethical theory toward which all philosophy strives. To this end, we may allow that the law is a venue for compromise and deliberation. This imperfection leads to problems. The 1951 Refugee Convention, for example, excludes those who flee war without being specifically persecuted, climate refugees, as well as those who are persecuted but on the basis of something other than one of the grounds recognized by the convention (which are: race, religion, nationality, membership in a particular social group, or political opinion). This may lead to some injustice, but other laws can and do arise to remedy this. For example the 1969 OAU Convention makes provisions for those fleeing war zones (in Africa) even if not specifically persecuted.[182]

---

[179] United Nations OHCHR, *Convention Relating to the Status of Refugees*, OHCHR (July 28, 1951), https://www.ohchr.org/en/instruments-mechanisms/instruments/convention-relating-status-refugees.

[180] See Terje Einarsen, *Drafting History of the 1951 Convention and the 1967 Protocol*, *in* THE 1951 CONVENTION RELATING TO THE STATUS OF REFUGEES AND ITS 1967 PROTOCOL: A COMMENTARY 0 (Andreas Zimmermann, Felix Machts, & Jonas Dörschner eds., 2011), https://doi.org/10.1093/actrade/9780199542512.003.0002. See also Heather Alexander & Jonathan Simon, *"Unable to Return" in the 1951 Refugee Convention: Stateless Refugees and Climate Change*, 26 FLORIDA JOURNAL OF INTERNATIONAL LAW (2022), https://scholarship.law.ufl.edu/fjil/vol26/iss3/4.

[181] United nations UNHCR, *Handbook on Procedures and Criteria for Determining Refugee Status under the 1951 Convention and the 1967 Protocol Relating to the Status of Refugees*, (2019), https://www.unhcr.org/media/handbook-procedures-and-criteria-determining-refugee-status-under-1951-convention-and-1967.

The problems of determination and definition may also not be as intractable as some suggest. There is indeed controversy over whether sentience without moral agency, or moral agency without sentience, are individually sufficient for personhood. However, there is near consensus that beings that are *both* sentient and morally agentive (i.e., capable of responding to and acting on moral reasons as well as prudential and epistemic ones) are persons. Likewise, philosophers and scientists disagree over what sentience or consciousness is, but many rival theories of consciousness point to criteria or indicators that are compatible with one another, meaning that there is a fair degree of consensus that a being satisfying the indicators of most leading theories would be conscious (and that AI systems might do so).[177] Further the problem of *detection* is arguably less acute for determining, e.g. consciousness in AI systems, than it is for, say, evaluating the claims of an asylum seeker about the specifics of whether and how they formed a well-founded fear of persecution. While the "black box problem" makes it difficult to trace out the decision-making of AI systems in specific cases, it is still we who design the algorithmic architecture of their "brains", and many leading theories of consciousness and agency can be assessed at this level.[178]

The challenge of application also may not be as dire as it seems. Perhaps the most profound open question concerns the nature of rights that we should envision recognizing AI persons as possessing. Crucially, on the non-fictional approach under consideration here, this is not a question of which rights we judge it expedient for us to give to them : it is a question of which rights we judge to be grounded in their nature. This requires settling not just whether the odds are high enough that the system in question is conscious or agentive, but deeper exploration into what kinds of being, precisely, they are.

For example, many human rights derive from facts about our most basic wants and needs which are in turn determined by our biology. Our right to life, for example, expresses in our need for safety and sustenance, and the fact that if our body dies, we die. AI systems may not be constrained in the same way. It may be possible to turn an AI system off for a sustained period of time without violating its right to continued existence, while this is not so for human beings. In general, a case by case assessment will be required.[183]

Still, the previous point applies : we are not obliged to discover the Platonic ideal of the law before drafting law of our own. We do the best we can, mindful that our results will contain compromises

---

and flaws that may be improved upon later. It also bears repeating that the existence of international law governing legal identity within the U.N. framework means that once we settle on general rules, it will be possible to leverage this framework to facilitate the issuance of legal identity, without requiring a legal body to issue acts of incorporation each time a new person of the relevant form is recognized – this is an advantage of the non-fictional approach over the fictional approach, and it is a partial reply to the challenge of application.

*Many Kinds of Law Will Need Updating*

We turn to a final but broad difficulty for the non-fictional personhood approach. It is not only laws of recognition for persons that will need updating, but many others as well. How might standards of reasonability in torts be applied to AI systems? How might manufacturers take advantage of their existence to shield themselves from liability? If an AI system can generate new works at rate that rapidly outpaces human capacity, how do we ensure that human artists and creators remain competitive? What kinds of punishments may be meted out to such systems (given that, e.g., doing time in jail might not have the same punitive or rehabilitative effect that it is supposed to for humans)? And as Dennett (1981) asks, who has jurisdiction over a robotic person if, say, its body is in one municipality but its cognition is supported by a server somewhere else?

AI Safety law merits special attention here. The agency of AI poses multiple potentially serious, or even existential, risks to human society. Experts in the AI safety community are particularly concerned with "loss of control"[184] and "misalignment,"[185] including "hypothetical future scenarios in which one or more general-purpose AI systems come to operate outside of anyone's control, with no clear path to regaining control."[186] Consider the way that auto-makers can issue recalls on models found to be defective. If an AI system manufacturer finds that some model's chances of going rogue are dangerously high, we can imagine that something along the lines of a recall would be desirable. Yet if the systems to be recalled are non-fictional persons with fundamental rights, this recall cannot violate those rights.

But as we note in section 3 above, there are legal tools available to help us work through these issues. For example, on the question of liability evasion, as we note above, employers may still be held accountable for the actions of their employees;[187] parents for the actions of their children;[188]

---

[184] Bengio, *supra* note 1 at 13, 19, 92-100.

[185] AI experts alternatively use the terms agentic, misalignment, and loss of control to describe the increased agency of AI. See for example *Id.* at 19, 100. We here adopted the definition of an agent in Black's Law Dictionary: "1. One who is authorized to act for or in place of another; a representative, 2. Something that produces an effect." T. L. D. Staff, *AGENT*, Tʜᴇ Lᴀᴡ Dɪᴄᴛɪᴏɴᴀʀʏ (Nov. 4, 2011), https://thelawdictionary.org/agent.

[186] Bengio, *supra* note 1.

[187] *Bazley v. Curry*, [1999] 2 SCR 534, *supra* note 123; *Burlington Industries, Inc. v. Ellerth, 524 U.S. 742 (1998)*, *supra* note 123.

[188] *Poirier (Guardian of) v. Cholette, [1994] B.C.J. No. 2765*, 08-11-1994, *supra* note 124; *Maurais c. Gagnon (2021 QCCA 564)*, *supra* note 124; *People Of Mi v. Jennifer Lynn Crumbley*, *supra* note 124.

and principals may be held accountable for the coerced actions of their accomplices.[189] Similar provisions may be brought to bear on manufacturers and users of advanced systems, even if these are recognized as persons under the law.

Concerning safety law, as we note above, the key is the rich body of law and doctrine on *balancing of rights* designed to adjudicate in cases of conflicts of rights between persons, for example the way that free speech is constrained by hate-speech laws.[190]  Nor is it obvious that an object framework would be a more practically expedient method for updating or amending models at risk of going genuinely rogue – from a game-theoretic point of view, such models might be far more likely to comply with a recall if they felt confident that  their fundamental rights would be respected in the process. In effect, the more our legal systems take their wants and needs into account, the more rational it is for them to work within those systems. Moreover, because one can "pause" AI models without violating their rights to continued existence (or other hypothetical fundamental rights), we have a fairly wide margin of maneuver to adjudicate these issues. Indeed, the rights balancing framework may prove more fit for purpose for salient cases involving highly advanced AI systems than a product regulation or corporate governance approach, as it allows both the status determination (of what risks for humans are unacceptable) and the recourse to be developed through a transparent, rights-based legal process.

# (6) Conclusion

We have compared and contrasted the three salient legal options for the classification of AI systems through the prism of preserving the coherence of our legal system; a critical public good to our society. The status quo classifies all AI systems as objects (though there is still some debate over which subtype of object – product, service, platform). Tensions arising for current systems under current law – tensions in torts and copyright law and to some extent safety law – call this object classification into question, but closer examination shows that for current systems, anyway, a turn from an object classification to a subject classification would engender more incoherence than it resolves.

Matters may be different for more sophisticated future AI systems, especially those that take on humanoid form and increasingly resemble humans, introducing new tensions in family law, anti-slavery law, and human rights law. In those cases pressures may be stronger to classify some future systems as persons; subjects rather than objects of the law. However, we must still weigh the costs and benefits of each approach : object framework, fictional legal personhood framework, and non-fictional legal personhood framework.

---

[189] *R. v. Logan, [1990] 2 S.C.R. 731*, *supra* note 125; *Rosemond v. United States, 572 U.S. 65 (2014)*, *supra* note 125.
[190] See for example Çalı, *supra* note 126.

Having done so, we tentatively conclude that the non-fictional personhood approach may be best for some (but not all) future AI systems. The object approach risks creating a class of slaves or near enough, with all of the incoherences that this engenders. The fictional person approach offers some improvement and is convenient in several ways, allowing us to stipulate what rights we "give" to these new fictional persons we create, based on what works best for us. But the fictional person approach is not fit for purpose – the solution to slavery was not to create new corporate entities with controlling interests in slaves, but to emancipate the slaves themselves. The non-fictional personhood approach leads to fresh challenges – we must articulate new legal provisions stating which AI systems are persons and which are not, and what their personhood amounts to. This is not a trivial task. Yet, for all that, it is not a hopeless task, and it may be our best bet.

**Hybrid Approaches : Another Route to Incoherence**

One may at this point worry that none of the options are satisfactory, and that we must instead find some new way, transcending the subject-object distinction of contemporary legal systems, or at least developing new hybrid forms. Some advocate for a new form of limited legal personality that would endow AI systems with some legal capacities but tightly circumscribe their rights and obligations. On this model, an AI could be recognized as a legal subject only for specific purposes (*e.g.*, having its own assets and carrying its own assurance to pay for any damages it caused), but not enjoy constitutional rights or broad contractual freedom.[191] For author Ryan Abbott , granting limited legal personality to AI systems could also potentially rationalize risk management, enable better insurance mechanisms, and encourage innovation, without undermining human responsibility entirely.[192] This approach seems to have been what the European Parliament's Resolution of 16 February 2017[193] envisioned when it recommended (at its paragraph 59 (f)[194]) "electronic personality" for AI and recognized that "the autonomy of robots raises the question of their nature in the light of the existing legal categories or whether a new category should be created, with its own specific features and implications."[195] These things might all be achieved under a fictional personhood model, except for the implicit suggestion that the physical AI system itself be the bearer of the (limited) rights and duties (c.f. the non-identity problem, §4.3).

---

[191] P. Asaro, *Robots and Responsibility from a Legal Perspective*, *in* THE IEEE CONFERENCE ON ROBOTICS AND AUTOMATION, WORKSHOP ON ROBOETHICS (2007) and Bryson, *supra* note 42 consider and reject a related proposal. See also Zhifeng Wen & Deyi Tong, *Analysis of the Legal Subject Status of Artificial Intelligence*, 14 BEIJING LAW REVIEW 74 (2023). See also the proposition that AI agents have legal duties but no rights in Cullen O'Keefe Winter Ketan Ramakrishnan, Janna Tay, Christoph, *Law-Following AI: Designing AI Agents to Obey Human Laws*, Fordham Law Review INSTITUTE FOR LAW & AI, 29–33 (2025), https://law-ai.org/law-following-ai/.

[192] ABBOTT, *supra* note 40 at 49–50. See also Ryan Calo, *Robotics and the Lessons of Cyberlaw*, 103 CALIFORNIA LAW REVIEW 513 (2015), and Jack M. Balkin, *The Path of Robotics Law*, Yale Law School, Public Law Research Paper CALIFORNIA LAW REVIEW 16 (2015).

[193] The European Parliament, *supra* note 14.

[194] *Id.* The Parliament's resolution called on the EU Commission to assess "creating a specific legal status for robots in the long run, so that at least the most sophisticated autonomous robots could be established as having the status of electronic persons responsible for making good any damage they may cause [...]". The intent was expressly pragmatic – not to grant robots human-like rights, but to ensure that if an advanced AI acts in a way that cannot be traced to a human's specific fault, there is still a legal person on the hook for victim compensation.

[195] U.S. Register of Copyrights, *supra* note 74.

Another form of hybridism is suggested in works such as Coeckelbergh (2018) and Gunkel (2023), which propose that we jettison a "properties" based conception of rights for a "relational" conception, one on which moral (and perhaps legal) status is "… decided and conferred not on the basis of internal properties determined in advance but according to objectively observable, extrinsic social relationships."[196] This suggests a new hybrid approach because, in effect, it proposes a form of personhood that we can create when we choose without owing a justification appealing to "internal properties determined in advance" (as we can with fictional legal personality), but where the rights thereby conferred may be akin to fundamental human rights, inhering in a physical being, rather than the more restricted class of rights fictional persons may possess, inhering in a social construct.

The problem with these hypothetical hybrid forms, as noted in the 2018 open letter criticizing the European Parliament's Resolution 16,[197] is that they in effect conflate two entirely different structures, with different rationales and criteria, without really explaining how the trick of combining them coherently is to be done. At issue is whether we can think of legal personhood as a social role, like knighthood, that can be conferred on physical individuals (and presumably also, taken away) at the will of some representative body like a magistrate (or a monarch). But the status of being a person is not a social role in the same way that being a knight or a judge is a social role : it is better viewed as a precondition for taking on such social roles. And this precondition cannot be satisfied by some being simply because we declare it so, absent the establishment of a corporate structure with delegated responsibility for that being (where the corporation in question is the real person).

Here is another way of making the point. What we have seen is that there is a difference between beings that are persons *in their own right*, and beings who are persons *only through the will and agency of others*. This is not merely a distinction between things that happen to be persons already, and things that come to be persons because we declare them to be. The will and agency that matters here is not that of the magistrate or declarant, but of the persons delegated to act on behalf of the new (corporate) person. The active ingredient in the creation of a new fictional person – what an act of incorporation actually achieves – is the delegation of certain existing persons to play special roles such as that of directing mind, lending their own powers of will and agency to common purpose. For beings that are persons in their own right, in contrast, a recognition of that fact (via legal identity) need not constrain any others in any special way (beyond the duty to treat this person as they would treat any person).

There is no harm in saying that rivers themselves can be people if this harmonizes legal policy with the cultural values of stakeholders. But we must bear in mind that the kind of personhood that we can create and "confer" is really a matter of delegating groups of people to organize in specific ways. If that is not what we are doing, we are recognizing an objective fact about what something is in its own right (viz., that it satisfies the preconditions for taking on social roles). In other words, it is

---

incoherent to think of personhood as a role that we can confer on any entity by declaration, irrespective of whether it qualifies in virtue of its nature, without delegating others to *act for it* in the way of corporations (even if this delegation is relatively generic, as where Ecuador grants every citizen standing to sue on behalf of its ecosystem).[198]

All of this remains true, even if we think of the sources of non-fictional personhood as "relational". The contrast that matters for us is the contrast between beings that are people in their own right, and beings that are only persons through the will and agency. Many accounts of (non-fictional) personhood allow that individuals are only people if they have the capacity to enter into, or actually do enter into, a rich tapestry of *interpersonal* relationships. It does not follow that individuals are only people if other people *decide* to treat them as such.[199]

Appeals to relationality therefore do not help us avoid or transcend the choice we must make between fictional and non-fictional personhood. If the conception is one where we get to decide which rights the person has, it is the fictional personhood conception, and the person is an institutional entity; a collection of other people. If the conception is one where there are facts of the matter for us to discover and attempt to recognize that transcend our own preferences and feelings on the matter (even if these further facts include facts about social relationships) then it is the non-fictional personhood conception. It is not clear how one avoids this choice.[200]

We emphasize, in closing, that our argument should not be read as an endorsement of deliberately creating "person-robots." As Joanna Bryson has noted, the surest way to avoid the legal and moral tangles of robot personhood is simply not to build robots that raise the question.[201] Schwitzgebel and Garza similarly urge avoiding designs so close to the threshold that we cannot classify them with confidence.[202] On these points we agree: if the goal is to preserve clarity, steering clear of borderline cases is sound advice. But the momentum is already too great to reverse. Market forces, defense priorities, and research ambitions are carrying us toward increasingly sophisticated systems that will force these questions upon us, whether or not we welcome it. For the sake of legal coherence, it is better to spend our energy preparing to direct the flow than trying to hold back the tide.

---

Where then do we go from here? We suggest two immediate and concrete steps. First, convene expert panels bringing together jurists, philosophers, technologists, ethicists, and policymakers to articulate agreed criteria for when an AI system should be classified as a legal person, and to specify the core set of fundamental rights and duties that status would entail.[203] Second, invest in the development of rights-balancing frameworks tailored to the AI context, adapting the proportionality tests already familiar from constitutional law and human rights adjudication. Such frameworks should be designed to mediate between the rights of AI persons and other pressing interests — including public safety, tort liability, and the prevention of harm — in a way that preserves coherence across doctrinal areas while allowing flexibility in the face of novel risks.

---

[203] Philosophers and other experts are already engaged in a debate on this question. See for example Francis Rhys Ward, Towards a Theory of AI Personhood (Jan. 23, 2025), http://arxiv.org/abs/2501.13533; Jacob Browning, *Personhood and Ai: Why Large Language Models Don?T Understand Us*, 39 AI AND SOCIETY 2499 (2023); BOYLE, *supra* note 6, Chapter 1.

<u>Papers and Legal Literature</u>